\documentclass[12pt]{article}
\pdfoutput=1  
\usepackage{amsmath}
\usepackage{amssymb,amsfonts}
\usepackage[all]{xy}
\usepackage{graphicx}
\usepackage[parfill]{parskip}
\usepackage[usenames]{color}
\usepackage{sidecap}
\usepackage{pdfpages}
\usepackage[utf8]{inputenc}

\numberwithin{equation}{section}
\numberwithin{table}{section}

\newcommand{\be}{\begin{equation}}
\newcommand{\ee}{\end{equation}}
\def\beq{\begin{eqnarray}}
\def\eeq{\end{eqnarray}}
\def\ba{\begin{eqnarray}}
\def\ea{\end{eqnarray}}

\def\ep1{\epsilon_1}
\def\eps2{\epsilon_2}

\newcommand{\IZ}{\mathbb{Z}}
\newcommand{\IC}{\mathbb{C}}
\newcommand{\IP}{\mathbb{P}}
\newcommand{\IN}{\mathbb{N}}
\newcommand{\IR}{\mathbb{R}}
\newcommand{\IQ}{\mathbb{Q}}

\newcommand{\im}{\mathrm{Im\, }}
\newcommand{\re}{\mathrm{Re\, }}

\newcommand{\arccosh}{\mathrm{arccosh}}

\newcommand{\tp}{\tilde{p}}
\newcommand{\tx}{\tilde{x}}

\newcommand{\bd}{{\boldsymbol{d}}}

\newcommand{\nn}{\nonumber}

\newcommand{\cN}{{\cal N}}
\newcommand{\cM}{{\cal M}}

\newcommand{\cO}{{\cal O}}
\newcommand{\cC}{{\cal C}}

\newcommand{\cF}{{\cal F}}

\newcommand{\tcO}{\tilde{\cO}}

\newcommand{\psiw}{\psi_{\mbox{\tiny WKB}}}
\newcommand{\psiwpm}{\psi_{\mbox{\tiny WKB}}^{\pm}}
\newcommand{\psiBPS}{\psi_{\mbox{\tiny BPS}}}

\newcommand{\Psiw}{\Psi_{\mbox{\tiny WKB}}}

\newcommand{\PsiBPS}{\Psi_{\mbox{\tiny BPS}}}

\newcommand{\Fns}{F_{\mbox{\tiny NS}}}

\newcommand{\Ftopopen}{F_{\mbox{\tiny top,open}}}
\newcommand{\FtopopenNS}{F_{\mbox{\tiny top,open}}^{\mbox{\tiny NS}}}
\newcommand{\FtopopenGW}{F_{\mbox{\tiny top,open}}^{\mbox{\tiny GW}}}
\newcommand{\FtopopenGWNS}{F_{\mbox{\tiny top,open}}^{\mbox{\tiny GW,NS}}}
\newcommand{\FtopopenBPS}{F_{\mbox{\tiny top,open}}^{\mbox{\tiny BPS}}}
\newcommand{\FtopopenLin}{F_{\mbox{\tiny top,open}}^{\mbox{\tiny lin}}}

\newcommand{\Ztopopen}{Z_{\mbox{\tiny top,open}}}
\newcommand{\ZtopopenNS}{Z_{\mbox{\tiny top,open}}^{\mbox{\tiny NS}}}
\newcommand{\ZtopopenGW}{Z_{\mbox{\tiny top,open}}^{\mbox{\tiny GW}}}
\newcommand{\ZtopopenBPS}{Z_{\mbox{\tiny top,open}}^{\mbox{\tiny BPS}}}
\newcommand{\ZtopopenBPSNS}{Z_{\mbox{\tiny top,open}}^{\mbox{\tiny BPS,NS}}}

\newcommand{\Ztop}{Z_{\mbox{\tiny top}}}

\newcommand{\Ftop}{F_{\mbox{\tiny top}}}
\newcommand{\FNS}{F_{\mbox{\tiny top,closed}}^{\mbox{\tiny NS}}}
\newcommand{\FNSpert}{F_{\mbox{\tiny top,closed}}^{\mbox{\tiny NS,pert}}}
\newcommand{\FNSBPS}{F_{\mbox{\tiny top,closed}}^{\mbox{\tiny NS,BPS}}}
\newcommand{\FNSBPSnp}{F_{\mbox{\tiny top,closed}}^{\mbox{\tiny NS,BPS,np}}}

\newcommand{\phipert}{\phi_{\mbox{\tiny pert}}}
\newcommand{\phiBPS}{\phi_{\mbox{\tiny BPS}}}

\newcommand{\V}{\Xi}
\newcommand{\lV}{\xi}
\newcommand{\vBPSp}{\zeta^{\mbox{\tiny BPS,+}}}
\newcommand{\vBPSn}{\zeta^{\mbox{\tiny BPS,-}}}
\newcommand{\vlin}{\zeta^{\mbox{\tiny const}}}

\newcommand{\Sodd}{S_{\mbox{\tiny odd}}}
\newcommand{\Seven}{S_{\mbox{\tiny even}}}

\newcommand{\functionspace}{\boldsymbol{F}_{x,\hbar}}

\newcommand{\mirrorcurve}{\boldsymbol{C}}
\newcommand{\cy}{\boldsymbol{X}}
\newcommand{\lambdaq}{\lambda_q}

\newcommand{\FS}{\cF}

\newcommand{\bQ}{\boldsymbol{Q}}
\newcommand{\bz}{\boldsymbol{z}}

\newcommand{\id}{\mathrm{id}}

\newcommand{\Res}{\mathrm{Res}}

\begin{document}
	
	\thispagestyle{empty}
	
\begin{flushright}LPTENS 16/02 \end{flushright}
	
	\vskip 3cm
	\noindent
	\begin{center}
		{\LARGE \bf Quantization condition from exact WKB\\for difference equations}
	\end{center}
	
	\vskip .1cm
	\linethickness{.05cm}
	\line(1,0){467}
	\vskip .5cm
	\noindent
	{\large \bf Amir-Kian Kashani-Poor}
	
	\noindent
	
	\vskip 0.15cm
	
	Laboratoire de Physique Th\'eorique de l'\'Ecole Normale Sup\'erieure,  \\
	Universit\'e de Recherche PSL,  CNRS,  Sorbonne Universit\'es, UPMC \\
	24 rue Lhomond, 75231 Paris Cedex 05, France

%
%
	
	\vskip 1cm

	\vskip0cm
	
	\noindent {\sc Abstract:} 
	A well-motivated conjecture states that the open topological string partition function on toric geometries in the Nekrasov-Shatashvili limit is annihilated by a difference operator called the quantum mirror curve. Recently, the complex structure variables parameterizing the curve, which play the role of eigenvalues for related operators, were conjectured to satisfy a quantization condition non-perturbative in the NS parameter $\hbar$. Here, we argue that this quantization condition arises from requiring single-valuedness of the partition function, combined with the requirement of smoothness in the parameter $\hbar$. To determine the monodromy of the partition function, we study the underlying difference equation in the framework of exact WKB.
	\vskip 1cm

	\pagebreak

	\tableofcontents

\section{Introduction}
A long standing goal of topological string theory is to obtain the topological string partition function $\Ztop$ as an analytic function on the parameter space of the theory. The latter is the product of the coupling constant space $\IC$ -- or $\IC^2$ in the case of refinement -- in which the genus counting parameter $g_s$ -- or $g_s$ and the coupling constant of the refinement $s$ -- take values, and the appropriate moduli space $\cM_{\cy}$ of the underlying Calabi-Yau manifold $\cy$. This program has been most successful in the case of toric (hence non-compact) Calabi-Yau manifolds. The topological vertex \cite{Aganagic:2003db} and its refined variants \cite{Iqbal:2007ii,Iqbal:2012mt} permit the computation of $\Ztop$ in the large radius regime of $\cM_{\cy}$ in a power series expansion in exponentiated flat coordinates of $\cM_{\cy}$, with coefficients that are rational functions in $e^{i\epsilon_{1}}$ and $e^{i\epsilon_2}$, with $g_s^2 = \epsilon_1 \epsilon_2$, $s= (\epsilon_1 + \epsilon_2)^2$. The holomorphic anomaly equations \cite{Bershadsky:1993cx} and their refinement \cite{Krefl:2010jb,Huang:2010kf,Krefl:2010fm, Huang:2011qx} can be used to compute the coefficients of $\Ftop = \log \Ztop$ in an asymptotic $(g_s,s)$ expansion as analytic functions on $\cM_{\cy}$. In the compact case, some impressive all genus results for certain directions in the K\"ahler cone have been obtained for Calabi-Yau manifolds that are elliptically fibered, see e.g. \cite{Huang:2015sta}. An open question is how to define $\Ztop$ without recourse to {\it any} expansion.

In \cite{Aganagic:2000gs,Aganagic:2001nx}, the open topological string partition function $\Ztopopen$ on a toric Calabi-Yau manifold $\cy$ was studied for a particular class of torically invariant branes, and the mirror curve $\mirrorcurve$ of $\cy$ identified as the open string moduli space for this problem. This insight led to the computation of $\Ftopopen$ to leading order in $g_s$. \cite{Aganagic:2003qj} proposed that to extend the computation beyond leading order in $g_s$, the mirror curve $\mirrorcurve$ had to be elevated to an operator $\cO_{\mirrorcurve}$. In fact, it is the Nekrasov-Shatashvili (NS) limit \cite{Nekrasov:2009rc} $g_s \rightarrow 0, s = \hbar^2$ of $\Ztopopen$ that can be determined via a quantization of the mirror curve, as $\ZtopopenNS$ lies in the kernel of $\cO_{\mirrorcurve}$ \cite{Nekrasov:2009rc,Mironov:2009uv,Aganagic:2011mi},\footnote{A different path towards such a quantization via the study of defects in five dimensional gauge theory is taken in \cite{Bullimore:2014awa}.}
\be \label{kernel_mirror_curve}
\cO_{\mirrorcurve} \ZtopopenNS = 0 \,.
\ee
The idea to recover the {\it closed} topological string partition function from the monodromy of the open partition function was put forward in \cite{Aganagic:2003qj}, and made more precise in \cite{Nekrasov:2009rc,Mironov:2009uv,Aganagic:2011mi}. In a remarkable series of papers \cite{Kallen:2013qla,Huang:2014eha,Grassi:2014zfa,Gu:2015pda,Wang:2015wdy,Codesido:2015dia,Hatsuda:2015qzx,Franco:2015rnr}, the quantization of the mirror curve has been taken as a framework within which to define $\Ztop$ non-perturbatively. In the genus one case, the equation (\ref{kernel_mirror_curve}) can straightforwardly be rewritten as a spectral problem for the complex structure parameter $z$ of the mirror geometry,
\be \label{spectral_problem_mirror_curve}
\tcO \ZtopopenNS= z \ZtopopenNS \,.
\ee
Here, $\cO_{\mirrorcurve}$ is put in the form $\tcO - z$ via appropriate variable redefinitions \cite{Codesido:2015dia}.\footnote{The significance of the choice of variables upon quantization of the mirror curve has been addressed in various works \cite{Aganagic:2003qj,Aganagic:2005dh,KashaniPoor:2006nc,KashaniPoor:2008xg}, but a complete understanding is still lacking.}
Upon specifying the function space $\FS$ in which $\ZtopopenNS$ is to lie, this eigenvalue problem can be solved numerically. For the higher genus case, \cite{Franco:2015rnr} identify the mirror curve $\mirrorcurve$ of the toric Calabi-Yau manifold $\cy$ with the spectral curve of a quantum integrable system determined by the toric data of $\cy$. The underlying class of quantum integrable systems was introduced by Goncharov and Kenyon \cite{Goncharov:2011hp}. The complex structure parameters $z_i$ of the mirror curve map to the spectrum of the integrable system. \cite{Kallen:2013qla} and follow-up works propose a quantization condition on the parameters $z_i$ based on a non-perturbative modification of $\FNS$, which roughly takes the form (see equation (\ref{quantization_condition}) below for the precise statement)
\be \label{crude_qc}
\partial_{T_k} \left( \FNSpert + \FNSBPS + \FNSBPSnp \right)= n_k + \frac{1}{2} \,, \quad n_k \in \IN \cup \{0\} \,.
\ee
Here, $\FNSpert$ and $\FNSBPS$ are the conventional perturbative and enumerative contribution to the closed topological string partition function in the NS limit. $\FNSBPSnp$ is a contribution included in the quantization condition to cancel poles of $\FNSBPS$ in $q=\exp{(i \hbar)}$. Condition (\ref{crude_qc}), at real values of $\hbar$, has been shown to reproduce the numerical results obtained by diagonalizing the Hamiltonians of the associated Goncharov-Kenyon system numerically in a harmonic oscillator eigenbasis of $L^2(\IR)$ to high precision.

In this paper, we aim to establish that the quantization condition (\ref{crude_qc}) arises upon imposing single-valuedness of the elements of the kernel of $\cO_{\mirrorcurve}$. To this end, we need to determine the monodromy of solutions to (\ref{kernel_mirror_curve}) as functions of the complex structure parameters $z_i$ on which $\cO_{\mirrorcurve}$ depends. We propose to do this in the framework of exact WKB analysis applied to difference equations. The WKB analysis of difference equations has received some treatment in the literature (see e.g. \cite{DingleMorgan}), however, to our knowledge, not in the form we require for our study. We thus attempt to generalize to difference equations the approach presented e.g. in \cite{MR2182990} to the transition behavior of WKB solutions of differential equations. We find that the transition behavior in the case of linear potentials can be studied in detail. Difference equations lack the rich transformation theory required to lift the analysis rigorously to general potentials \cite{MR2182990}. Our analysis hence relies on the conjecture that the transition behavior for potentials with simple turning points is governed only by these, and well approximated in their vicinity by the linear analysis.

To explain the non-perturbative contribution to (\ref{crude_qc}), we will argue that by the choice of harmonic oscillator states for the numerical diagonalization, the elements of the function space $\FS$ are constrained to be $L^2$ functions in $x$ with smooth dependence on $\hbar$. The latter condition requires adding a non-perturbative piece in $\hbar$ to $\ZtopopenNS$ as hitherto defined. Equation (\ref{crude_qc}) arises from the constraint that the function thus obtained be single-valued.

We will study the monodromy problem of difference equations in section \ref{section:monodromy_from_WKB}. In section \ref{section:open_top_string}, we discuss the open topological string partition function from various perspectives and explain how we expect the quantization condition (\ref{crude_qc}) to arise. This discussion is applied to the example $\cO(K) \rightarrow \IP^1 \times \IP^1$ in section \ref{section:example}, in which we also present numerical evidence for the quantization condition (\ref{crude_qc}) in the case of complex $\hbar$. We end with conclusions.

\section{Monodromy from exact WKB} \label{section:monodromy_from_WKB}
In this section, we will provide evidence linking the monodromy problem of solutions to difference equations of the form (\ref{kernel_mirror_curve}) arising upon quantization of mirror curves to the so-called quantum B-periods. We first flesh out an argument provided by Dunham \cite{Dunham:1932} in the case of differential equations using exact WKB methods. We then set out to generalize these methods to difference equations.

\subsection{Differential equation} \label{subsection:ode}
We will briefly review the basics of WKB analysis in this subsection, following \cite{MR2182990}. A somewhat more detailed review in the same spirit can be found in \cite{Kashani-Poor:2015pca}.

The starting point of the analysis is a second order differential equation
\be \label{wkb_ode}
\left( \epsilon^2 \partial_x^2 - Q(x) \right) \Psi(x) = 0 \,,
\ee
depending on a small parameter $\epsilon$. $Q(x)$ is a meromorphic function of $x$ with possible $\epsilon$ dependence, which we for simplicity will take to be of the form $Q(x) = \sum_{n=0}^N Q_{2n}(x) \epsilon^{2n}$. To solve this equation, we can make a WKB ansatz
\be \label{wkb_ansatz_ode}
\psiw(x) = \exp \int^x S(x) \,dx \,,
\ee
with $S$ considered as a formal power series in $\epsilon$,
\be  \label{WKB_differential}
S(x)= \frac{1}{\epsilon}S_{-1}(x) + \sum_{n=0}^\infty S_n(x) \epsilon^n \,.
\ee
Plugging this ansatz into the differential equation (\ref{wkb_ode}) yields the expansion coefficients $S_n$ recursively,
\ba
S_{-1}^2(x) &=& Q_0(x) \,, \label{leading_ode}\\
2 S_{-1} S_{n+1} + \sum_{n_1 =0}^n S_{n}(x) S_{n-n_1}(x) + \frac{S_n(x)}{dx} &=& Q_{n+2} \,, \quad n>-1 \,.
\ea
The equation (\ref{leading_ode}) has two solutions $S_{-1} = \pm \sqrt{Q_0(x)}$. The choice of sign propagates down to all expansion coefficients $S_{2n+1}$. We thus obtain two formal WKB solutions $\psiwpm$ to (\ref{wkb_ode}), reflecting the fact that the differential equation is of second order.

Denoting 
\be
\Sodd = \sum_{n \, \mbox{\tiny odd}} S_n \epsilon^n \,, \quad \Seven = \sum_{n \, \mbox{\tiny even}} S_n \epsilon^n \,,
\ee
it is not hard to show that
\be
\Seven = -\frac{1}{2} \frac{d \log \Sodd}{dx} \,.
\ee
The two formal WKB solutions can thus be expressed as
\be \label{gen_sol_ode}
\psiwpm(x) = \frac{1}{\sqrt{\Sodd}} \exp\left( \pm \int^x \Sodd \, dx \right) \,.
\ee
The two formal series $\psiwpm$ will generically merely provide asymptotic expansions of two solutions to (\ref{wkb_ode}). Exact WKB analysis is concerned with recovering the functions underlying such expansions.

Borel resummation is a technique to construct a function having a given asymptotic expansion as a power series
\be \label{formal_exp}
\psi(\epsilon) = \sum_{k=0}^\infty \psi_k \epsilon^k \,.
\ee
It proceeds in two steps. The first is to improve the convergence behavior of (\ref{formal_exp}) by considering the Borel transform
\be \label{Borel_transform}
\psi_B(y) = \sum_{k=1}^\infty \psi_k \frac{y^{k-1}}{(k-1)!} \,.
\ee
The second step is to take the Laplace transform of (\ref{Borel_transform}),
\be \label{Laplace_transform}
S_{\theta}[\psi](\epsilon) = \psi_0 + \int_{\ell_\theta} e^{-\frac{y}{\epsilon}} \psi_B(y) \, dy \,.
\ee
Here, $\ell_\theta$ is a half-line in the $y$-plane, emanating from the origin at angle $\theta$ to the abscissa. If the sum in (\ref{Borel_transform}) and the integral in (\ref{Laplace_transform}) exist, then $S_{\theta}[\psi](\epsilon)$ defines a function with asymptotic expansion given by (\ref{formal_exp}), called a Borel resummation of the formal power series (\ref{formal_exp}). The Laplace transform (\ref{Laplace_transform}) can fail to exist if $\psi_B(y)$ exhibits a singularity on the integration path $\ell_\theta$. Integrating along $\ell_{\theta^\pm}$ on either side of the singularity will then generically give rise to two functions $S_{\theta^\pm}[\psi](\epsilon)$, both with asymptotic expansion (\ref{formal_exp}), but differing by exponentially suppressed pieces in $\frac{1}{\epsilon}$. The position of the singularities of the Borel transform $\psi_B$ thus leads to a subdivision of the $y$-plane into sectors. Choosing $\ell_\theta$ to lie in different sectors will give rise to different Borel resummations of (\ref{formal_exp}).

When the coefficients of the formal power series (\ref{formal_exp}) depend on a variable $x$, 
\be \label{formal_exp_x}
\psi(x,\epsilon) = \sum_{k=0}^\infty \psi_k(x) \epsilon^k \,,
\ee
the position of the poles of the Borel transform (\ref{Borel_transform}), and hence the subdivision of the $y$-plane into sectors, will depend on $x$. Keeping the integration path $\ell_\theta$ fixed, crossing certain lines in the $x$-plane will result in poles of $\psi_B$ crossing $\ell_\theta$. These lines are called Stokes lines. They divide the $x$-plane into Stokes regions. The Borel resummation of (\ref{formal_exp_x}) performed on either side of a Stokes line will yield functions whose analytic continuation to a mutual domain will differ by exponentially suppressed terms.

Returning to the WKB analysis of a second order differential equation, the Stokes phenomenon implies that a Borel resummation of the formal WKB solutions (\ref{gen_sol_ode}) will yield a different basis of the solution space depending on the Stokes region in which the Borel resummation is performed. This behavior can be studied by first considering the case of a linear potential $Q(x)=x$, then using the transformation theory of differential equations to reduce the analysis of more general potentials to this case. For potentials with only simple zeros, the results are as follows: the Stokes lines emanate from zeros of $Q_0$, called turning points. Simple turning points have three Stokes lines and a branch cut emanating from them. The trajectory of Stokes lines depends on the choice of integration path $\ell_\theta$ for the Laplace transform and is determined by the equation
\be \label{location_stokes_line}
\im e^{i \theta} \int_{x_0}^x S_{-1} \, dx  = 0 \,,
\ee
with $x_0$ the position of the turning point. As the Borel resummation of the formal WKB solutions $\psiwpm$ in any Stokes region yields a basis of solutions to the differential equation (\ref{wkb_ode}), each such pair can be expressed as a linear combination of any other such pair. Neighboring Stokes regions are assigned transition matrices which enact the linear transformation relating the associated two pairs of solutions. The form of the transition matrices depends on the normalization of the WKB solutions, determined by the lower bound on the integration in the exponential of the WKB ansatz (\ref{wkb_ansatz_ode}). Choosing this lower bound to be the turning point from which the Stokes line separating the two Stokes regions emanates yields $\epsilon$ independent transition matrices. 

The exact form of the transformation matrices can be determined, as mentioned above, by solving the differential equation with linear potential explicitly, and then mapping the general situation to this case. The space of solutions to the linear problem is spanned by Airy functions. For our purposes, we will only need the product $T$ of the three transition matrices which arise when we circumnavigate a turning point in counter clockwise order, crossing three Stokes lines consecutively, but without crossing the branch cut. To compute $T$, it suffices to know that the Airy functions are single-valued in the vicinity of the turning point. It follows that 
\be 
B \cdot T = \id \,,
\ee
where the matrix $B$ relates the Borel resummation of $\psiwpm$ in the same Stokes region, but on either side of the branch cut. Crossing the branch cut interchanges $\psiw^+$ and $\psiw^-$, and leads to a factor of $i$ due to the square root in the denominator of (\ref{gen_sol_ode}). This reasoning yields
\be \label{transition}
T = \pm i \begin{pmatrix}
	0 & 1 \\
	1 & 0 
\end{pmatrix} \,.
\ee
The sign depends on conventions that we will not bother to fix, as it will cancel in our considerations.

\begin{figure}
	\centering
	\includegraphics[width=5cm]{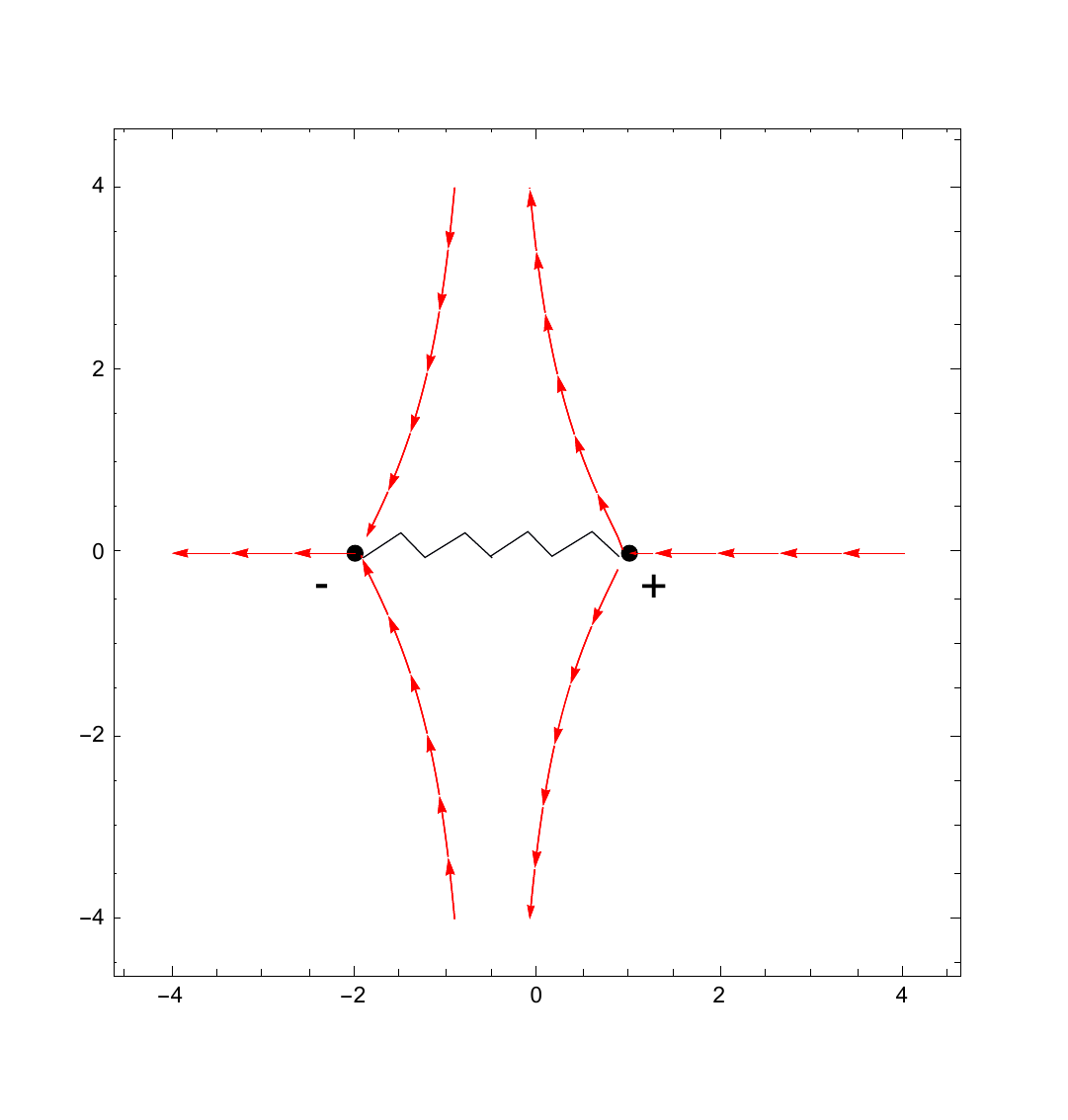}
	\caption{\small A generic potential with two turning points; the branch cut is chosen to connect the two turning points.} \label{figure_generic_pot}
\end{figure}

Let us now consider a potential with two simple turning points, giving rise to a Stokes pattern as depicted in figure \ref{figure_generic_pot}. 
We want to consider the monodromy of a pair of WKB solution as we encircle the two turning points once. If we begin with a pair of WKB solutions normalized at the turning point $x_1$, we must change their normalization by multiplying by the matrix
\be
N^{\mbox{\tiny SR}}_{1\rightarrow 2} =  \begin{pmatrix}
	\exp[\int_{x_2}^{x_1} \Sodd \,dx] & 0 \\
	0 & \exp[-\int_{x_2}^{x_1} \Sodd \,dx] 
\end{pmatrix}
\ee
in order for their transition behavior upon circumnavigating the turning point $x_2$ to be governed by (\ref{transition}). The superscript ${}^{\mbox{\tiny SR}}$ is to denote the Stokes region in which the integration path from $x_1$ to $x_2$ lies. The exponential entries in the normalization matrix are called Voros multipliers. They are to be understood as the Borel resummation of the indicated formal power series. Such Borel resummations exhibit interesting jumping behavior with regard to the choice of integration path $\ell_\theta$ for the Laplace transform, as e.g. recently discussed in \cite{Kashani-Poor:2015pca}. 

In total, we obtain the monodromy matrix
\be
N^{\mbox{\tiny SR}_a}_{2\,\rightarrow\,1}T N^{\mbox{\tiny SR}_b}_{1\,\rightarrow\,2} T = 
-\begin{pmatrix}
	\exp[\oint \Sodd \,dx] & 0 \\
	0 & \exp[-\oint \Sodd \,dx] 
\end{pmatrix} \,,
\ee
with the superscripts ${}^{\mbox{\tiny SR}_a}$ and ${}^{\mbox{\tiny SR}_b}$ indicating that the integration path connecting $x_1$ and $x_2$ is to be taken above/below the branch cut. The integration cycle is accordingly a path encircling the branch cut. 

Requiring a pair of WKB solutions in the situation depicted in figure \ref{figure_generic_pot} to be single-valued is hence equivalent to demanding
\be  \label{single_valuedness_condition}
\oint \Sodd \,dx = 2\pi i (n +\frac{1}{2})  \,.
\ee

\subsection{Difference equation} \label{subsection:diff_equ}
Unlike differential equations, difference equations have no obvious transformation theory. Under variable transformation other than linear, their form changes drastically. We will perform an exact WKB analysis in the case of linear potential in the following, but not be able to offer an intrinsic criterion determining for which potentials the linear approximation is justified.

\subsubsection{Generalities} \label{section:wkb_gen}
Consider a difference equation in the form
\be \label{gen_pot}
\big[\cosh \epsilon \partial_x - Q(x)\big] \Psi(x) = 0  \,,
\ee
for a potential $Q(x)$ which we take to be $\epsilon$ independent for simplicity. With the WKB ansatz
\be  \label{WKB_ansatz_difference}
\psiw(x) = \exp  \int^x S(x) \,dx = N(\epsilon) \exp R(x)\,,
\ee
$N(\epsilon)$ being a normalization factor which we shall fix below, we obtain
\be
\exp \left[ \sum_{n=1}^\infty \frac{1}{(2n)!} S^{(2n-1)}(x) \epsilon^{2n} \right]\cosh \left[ \sum_{n=0}^\infty \frac{1}{(2n+1)!}S^{(2n)}(x) \epsilon^{2n+1} \right] = Q(x) \,.
\ee
Expanding
\be  \label{WKB_difference}
S= \frac{1}{\epsilon}S_{-1} + \sum_{n=0}^\infty S_n \epsilon^n \,,\quad R= \frac{1}{\epsilon}R_{-1} + \sum_{n=0}^\infty R_n \epsilon^n \,,
\ee
we obtain
\be  \label{lambda}
\cosh S_{-1} = Q(x) \quad \Leftrightarrow \quad S_{-1} = \pm \arccosh\, Q(x) + 2 \pi i n \,, \quad n\in \IZ 
\ee
and furthermore
\ba \label{S0}
S_0 &=& -\frac{1}{2} S_{-1}' \coth S_{-1} = -\frac{1}{2} \frac{d}{dx} \log \sinh S_{-1} \\
&=& \frac{Q(x) Q'(x)}{2(1-Q(x)^2)} = -\frac{1}{4} \frac{d}{dx} \log (Q^2(x)-1) \,, \nn \\
S_1 &=& \pm \ldots \,. \nn
\ea
Hence,
\be \label{difference_wkb}
\psiwpm(x) = \frac{1}{(Q^2(x)-1)^{\frac{1}{4}}} \exp \left[ \pm \frac{1}{\epsilon}\int^x \arccosh \, Q(x) \,dx + \cO(\epsilon) \right] \,,
\ee
where we have chosen a branch in (\ref{lambda}).

The analytic structure of the inverse $\cosh$ function is best understood by expressing it as
\be \label{arccosh_as_log}
\arccosh\,z = \log (z + \sqrt{z-1}\sqrt{z+1})  \,.
\ee
The $\arccosh$ function has two branch points at $\pm 1$ respectively due to the square root functions, and one at $- \infty$ due to the logarithm.
Choosing the branch cuts for the square roots and the logarithm in the negative real direction, the branch cuts of the two square roots cancel beyond $z = -1$, at which point the branch cut of the logarithm begins.\footnote{The preimage of the negative real axis under
	\be
	z \mapsto z + \sqrt{z^2 -1}
	\ee
is the interval $(- \infty, -1]$.} The sheet structure of $\arccosh$ is hence such that the branch cut along $[-1,1]$ connects two sheets related via a sign flip, as
\be
\log (z + \sqrt{z-1}\sqrt{z+1})  \mapsto \log (z - \sqrt{z-1}\sqrt{z+1}) = -\log (z + \sqrt{z-1}\sqrt{z+1})  \,,
\ee
whereas the branch cut beyond $z=-1$ connects sheets related via a shift of the imaginary part by $2 \pi$. 

The zero of (\ref{arccosh_as_log}) lies at $z=1$. The expansion of $\arccosh z$ around this point has $\sqrt{z}$ as leading term. We thus identify the points $\{x_0: Q(x_0)=1\}$ as the turning points of the difference equation. As long as $Q'(x_0) \neq 0$, we can approximate the behavior of the difference equation in the vicinity of such a turning point by a linear potential.

\subsubsection{WKB for a linear potential} \label{subsection:linear_potential}
The difference equation with a linear potential is
\be \label{diff_equ_lin_pot}
\big(\cosh \epsilon \partial_x - x\big) \Psi(x) = 0  \,.
\ee
Setting $Q(x) = x$ in (\ref{lambda}) and (\ref{S0}), the WKB coefficients $S_{n}$ can be integrated, yielding
\begin{alignat}{2}
R_{-1}&=\pm \int S_{-1}(x) \,dx &&= \pm \left( \sqrt{x^2-1} - x\, \arccosh (x) \right)  \,, \label{S_diff_eq} \\
R_{0} &=\int S_{0}(x) \,dx &&= -\frac{1}{4} \log (-1 + x^2 ) \,,\quad \mbox{etc.} \nn
\end{alignat}
By (\ref{difference_wkb}) and using (\ref{arccosh_as_log}), the leading order behavior of the WKB solution is thus
\be \label{leading_wkb_lin}
\psiwpm(x) = \frac{N(\epsilon)}{(x^2-1)^{\frac{1}{4}}} \exp \left[\pm \frac{1}{\epsilon} \left(\sqrt{x^2-1} - x \log \left(x+\sqrt{x^2-1} \right) \right) + \cO(\epsilon) \right] \,.
\ee
In fact, we can check explicitly that this WKB solution provides an asymptotic expansion for a solution of the difference equation (\ref{diff_equ_lin_pot}), as we can construct a solution to this equation based on the Bessel function \cite{Ehrhardt}.\footnote{Bessel functions appear in \cite{Aoki:2016} in the analysis of Stokes curves of loop type. It would be interesting to explore connections to the analysis presented here.} Recall that the Bessel function satisfies the recursion relation
\be \label{recur_Bessel}
J_{\nu+1}(z) - 2 \frac{\nu}{z} J_\nu(z) + J_{\nu-1}(z) = 0 \,.
\ee
The function $ J_{\frac{x}{\epsilon}}(\frac{1}{\epsilon})$ thus solves the difference equation (\ref{diff_equ_lin_pot}).\footnote{This can also be seen immediately by considering the Fourier transform of the difference equation (\ref{diff_equ_lin_pot}). We thank Jorge Russo for this remark.}

The Bessel function is known to have asymptotic behavior, for $\nu \rightarrow \infty$ along the real axis and constant positive $z$, given by (see e.g. \cite{OLBC10})
\be \label{asymp_bessel}
J_\nu(\nu z) \sim \frac{1}{(2 \pi \nu \sqrt{1-z^2})^\frac{1}{2}} \exp \left[ \nu \left(\sqrt{1-z^2}- \log(1+\sqrt{1-z^2}) + \log z \right) \right] \,.
\ee
Upon the identification
\be
\nu \mapsto \frac{x}{\epsilon} \,, \quad z \mapsto \frac{1}{x} \,,
\ee
this coincides with the WKB result $\psiw^+$  (\ref{leading_wkb_lin}), with the normalization $N(\epsilon)$ fixed at
\be 
N(\epsilon) = \sqrt{\frac{\epsilon}{2 \pi}}  \,.
\ee
We have hence matched the leading behavior of the WKB solution (\ref{leading_wkb_lin}) to the asymptotic expansion of the Bessel function $J_{\frac{x}{\epsilon}}(\frac{1}{\epsilon})$ for positive real $x$ and $\epsilon$ small and positive. To study the Stokes phenomenon, we will now take advantage of the fact that the Bessel function is also the solution to a differential equation, to which we can apply the exact WKB methods reviewed in the previous section. Indeed, $J_{\nu}(z)$ solves the differential equation
\be
z^2 \frac{d^2}{dz^2} y + z \frac{d}{dz}y + (z^2 - \nu^2) y = 0 \,.
\ee
We can eliminate the linear term and cast the equation in the form (\ref{wkb_ode}) by considering $w_{\nu}(z) = \sqrt{z}\, y (\nu z)$, which satisfies
\be \label{Bessel_ode}
\frac{d^2}{dz^2} w_{\nu}(z) - \left( \frac{1-z^2}{z^2}\nu^2 - \frac{1}{4 z^2} \right) w_{\nu}(z) = 0 \,.
\ee
The conventional theory of exact WKB analysis of differential equations allows us to determine the Stokes behavior of the WKB expansion of the solutions to this differential equation. By relating this expansion to the WKB solution of the difference equation, we can derive the Stokes behavior of the latter.

Making a WKB ansatz\footnote{The superscript ${}^\partial$ is to distinguish quantities pertaining to the differential equation (\ref{Bessel_ode}) from those pertaining to the difference equation (\ref{diff_equ_lin_pot}).} $w \sim \exp R^\partial$, we obtain the leading terms
\ba
S^\partial_{-1} &=& \pm \sqrt \frac{1-z^2}{z^2} \,, \\
S^\partial_0 &=& - \frac{(S^\partial_{-1})'}{2S^\partial_{-1}} = - \frac{1}{2} \frac{d}{dz} \log S^\partial_{-1} \,.
\ea
By comparing to the asymptotic expansion (\ref{asymp_bessel}), we can fix the normalization of the WKB expansion to
\be \label{WKB_diff_normalized}
\sqrt{z} J_\nu (\nu z) \sim \frac{1}{\sqrt{2\pi\nu}} \exp \sum_{n=-1}^\infty R_n^\partial (z) \nu^{-n} \,.
\ee
Comparing to 
\be \label{WKB_difference_normalized}
J_{\frac{x}{\epsilon}} (\frac{1}{\epsilon}) \sim \sqrt{\frac{\epsilon}{2\pi}} \exp \sum_{n=-1}^\infty R_n(x) \epsilon^{n} \,,
\ee
the uniqueness of asymptotic expansions in power series implies
\be \label{relating_WKBs}
\sum_{n=-1}^\infty R_n(x) \epsilon^{n} =  \sum_{n=-1}^\infty R_n^\partial (\frac{1}{x}) \left(\frac{\epsilon}{x}\right)^{n} \,.
\ee
We have here assumed that the WKB expansions (\ref{WKB_difference_normalized}) and (\ref{WKB_diff_normalized}) yield asymptotic expansions to the indicated solutions of the difference and differential equation respectively. In the case of the differential equation, this is guaranteed by general theory. We have verified (\ref{relating_WKBs}) to high order in $\epsilon$.

\begin{figure}
	\centering
	\includegraphics[width=7cm]{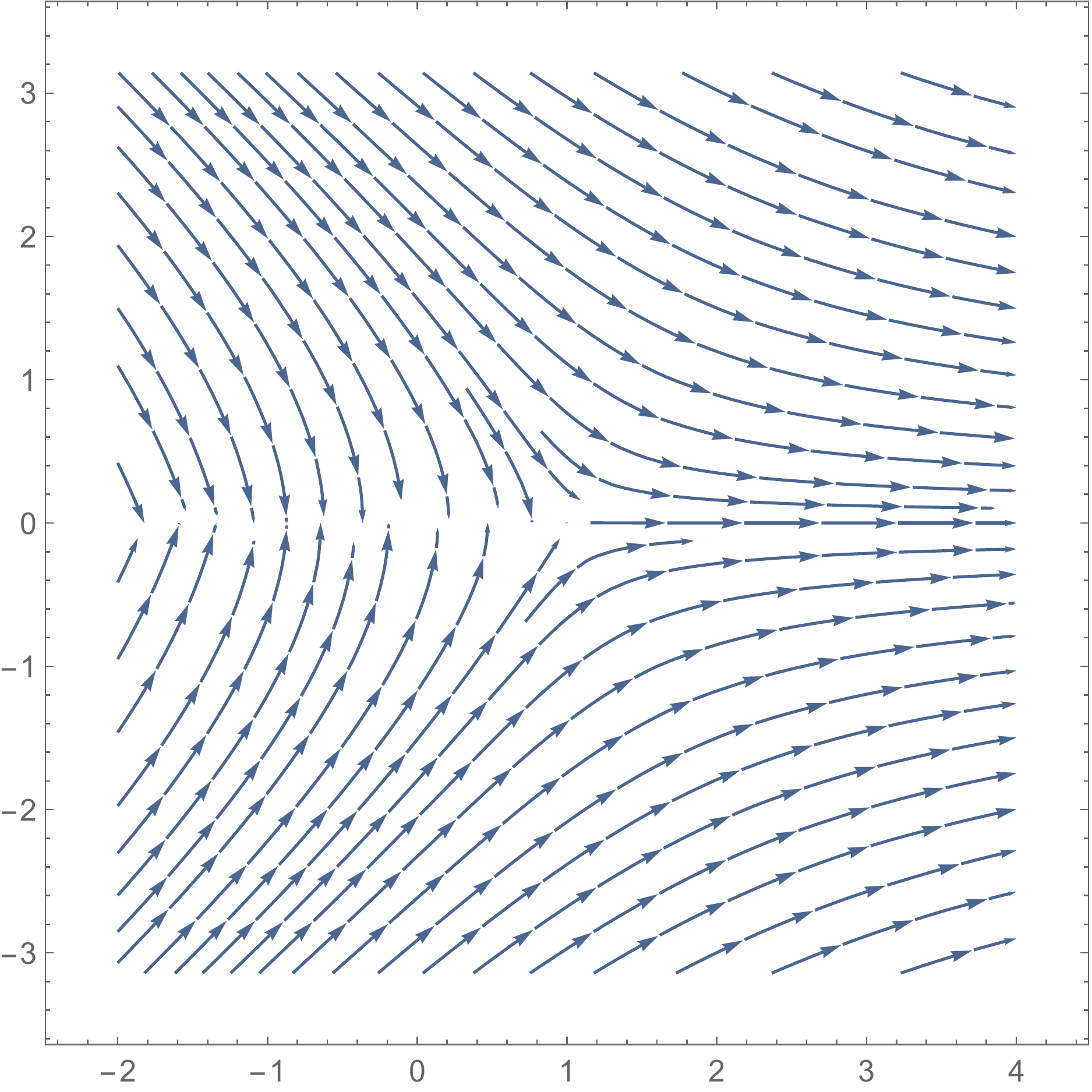}
	\caption{\small Flow lines for $S_{-1} = \arccosh(z)$.} \label{figure_flow_arccoshz}
\end{figure}

We next address the question of how the Stokes lines of the two asymptotic expansions 
\begin{alignat}{1}
\psiw^{\partial}(z)=  e^{\frac{1}{\epsilon} R_{-1}^\partial(z)} \sum_{k=0}^\infty \psi^\partial_k(z) \epsilon^{k+\frac{1}{2}} \quad \mbox{and} \quad\psiw(x) &= e^{\frac{1}{\epsilon} R_{-1}(x)} \sum_{k=0}^\infty \psi_k(x) \epsilon^{k+\frac{1}{2}} \\
&= e^{\frac{1}{\epsilon} x R^\partial_{-1}(1/x)} \sum_{k=0}^\infty \frac{\psi^\partial_k(1/x)}{x^k} \epsilon^{k+\frac{1}{2}} \nn
\end{alignat}
are related. The Borel transforms of the two expansions are given by
\be
\psi_B^{\partial}(z,y) = \sum_{k=0}^\infty  \frac{\psi^\partial_k(z)}{\Gamma(k+\frac{1}{2})} (y+R^{\partial}_{-1}(z))^{k-\frac{1}{2}}
\ee
and
\ba \label{relation_borel_transforms}
\psi_B(x,y) &=& \sum_{k=0}^\infty  \frac{\psi_k(x)}{\Gamma(k+\frac{1}{2})} (y+R_{-1}(x))^{k-\frac{1}{2}} \\
&=& \frac{1}{\sqrt{x}} \sum_{k=0}^\infty  \frac{\psi^\partial_k(\frac{1}{x})}{\Gamma(k+\frac{1}{2})} \left( \frac{y}{x}+R^\partial_{-1}(\frac{1}{x})\right)^{k-\frac{1}{2}} \\
& =& \frac{1}{\sqrt{x}} \psi_B^{\partial}(\frac{1}{x},\frac{y}{x} ) \,.
\ea
Hence,
\ba
\int_{-R_{-1}(x)}^{\infty} e^{-\frac{y}{\epsilon}} \psi_B(x,y) \, dy &=& \int_{-R_{-1}(x)}^{\infty} e^{-\frac{y}{\epsilon}}  \frac{1}{\sqrt{x}} \psi_B^{\partial}(\frac{1}{x},\frac{y}{x}) \, dy  \\
&=& \sqrt{x}\int_{-\frac{R_{-1}(x)}{x}}^{\infty} e^{-\frac{y}{\epsilon/x}}  \psi_B^{\partial}(\frac{1}{x},y) \, dy  \\
&=&  \sqrt{x} \,\psi^\partial(\frac{1}{x},\frac{\epsilon}{x}) \,.
\ea
The Borel sum of the WKB series of the difference equation hence indeed equals, for real $x$ and small positive $\epsilon$, $\sqrt{x} \psiw^\partial(\frac{1}{x},\frac{\epsilon}{x}) = J_{\frac{x}{\epsilon}}(\frac{1}{\epsilon})$. From the theory of exact WKB for differential equations, we know that the Borel transform $\psi_B^\partial(z,y)$ has a branch point in the $y$-plane at $R_{-1}^\partial(z)$. The Laplace transform performed along the real axis, i.e. with $\ell_\theta = \IR^+$ in the notation of (\ref{Laplace_transform}), will hence be ill-defined for $R_{-1}^\partial(z) \in \IR^+$, identifying this condition as determining the location of the Stokes line. By (\ref{relation_borel_transforms}), $\psi_B(x,y)$ hence exhibits a branch point at $y/x = R_{-1}^\partial(\frac{1}{x})$, i.e. $y = R_{-1}(x)$. The condition determining the location of the Stokes line is therefore $R_{-1}(z) \in \IR^+$. We conclude that the location of the Stokes lines of the difference equation is determined by the phase of $R_{-1}(z)$, just as a naive generalization of the conventional WKB results would have suggested. By
\be
R_{-1}(z) = \frac{-2 \sqrt{2}}{3}(z-1)^{\frac{3}{2}} + {\cal O}((z-1)^{\frac{5}{2}})  \,,
\ee
the Stokes line structure close to the turning point at $z=1$ is the same as around a simple turning point in the case of a differential equation of the form (\ref{wkb_ode}), see figure \ref{figure_flow_arccoshz}.

The behavior of the Borel resummed WKB solution $\Psiw^\partial$ upon crossing Stokes lines emanating from the turning point $z=1$ is governed by the general theory. In particular, the transition behavior of $\Psiw^\partial$ upon circumnavigating a turning point is given by the matrix (\ref{transition}), and $\Psiw$ inherits this behavior.

If we assume that the monodromy of the WKB solutions of difference equations is governed by Stokes lines emanating from turning points $\{x_0: Q(x_0)=1\}$, and that the behavior upon crossing such lines is captured by the analysis for linear potential just presented, then the analysis of section \ref{subsection:ode} applies, leading to the single-valuedness condition (\ref{single_valuedness_condition}) in the case of potentials with two turning points.

\section{The open topological string and the mirror curve} \label{section:open_top_string}
\subsection{The conjectured quantization condition} \label{subsection:qc}
We begin this section by reviewing the quantization condition discussed in the introduction as presented in \cite{Franco:2015rnr}. In this form, it applies to the topological string on an arbitrary toric Calabi-Yau manifold $\cy$. The mirror to such a space is given by a pair $(\mirrorcurve,\lambda)$, consisting of a complex curve $\mirrorcurve$ together with a meromorphic 1-form $\lambda$, the 5d analogue of the Seiberg-Witten differential \cite{Seiberg:1994rs}. $\mirrorcurve$ is given as the zero locus of a polynomial
\be \label{mirror_curve}
P_{\mirrorcurve}(e^x,e^p,e^{-x},e^{-p}) = 0 
\ee
which can be constructed, up to linear redefinitions of the variables $x$ and $p$, from the toric data of $\cy$ \cite{Chiang:1999tz}. The latter can be presented as a grid diagram, given by the intersection of the three dimensional fan of $\cy$ with the $x_3=1$ plane. The number of interior points of the grid diagram corresponds to the genus $g$ of $\mirrorcurve$. Each such point gives rise to a modulus $z_i$ which enters as a parameter in $P_{\mirrorcurve}$. Each of the $N$ boundary points of the grid diagram beyond the first three gives rise to an additional parameter $m_i$ or $z_{g+i}$ in $P_{\mirrorcurve}$, referred to as a mass parameter in \cite{Huang:2014nwa}.\footnote{These do not necessarily correspond to physical masses in the geometrically engineered theory; the geometry engineering pure $SU(2)$ e.g. exhibits such a mass parameter, and it corresponds to the scale $\Lambda$ at which the gauge theory is defined.}  The moduli $z_i$, $i=1, \ldots, g$, coincide in the large radius limit with $Q_i=\exp(-T_i)$, the exponentials of the flat coordinates $T_i$ on the complexified K\"ahler moduli space of $\cy$. These are chosen among the $g+N-3$ simply logarithmic solutions of the underlying Picard-Fuchs equations governing the periods of the meromorphic 1-form $\lambda$ on the curve $\mirrorcurve$. They are paired with doubly logarithmic solutions corresponding to $B$-periods. In contrast, the mass parameters $m_i$ or $z_{g+i}$ correspond to residues of the 1-form and do not have dual partners. They are given as algebraic functions of the $g+N-3$ exponentiated logarithmic solutions $Q_i$ to the Picard-Fuchs system.

The polynomial $P_{\mirrorcurve}$ can be promoted to an operator $\cO_{\mirrorcurve}$ by setting
\be \label{canonical_comm}
p  = \frac{\hbar}{i} \,\partial_x \,,\mbox{ such that} \quad [x,p] = i \hbar \,.
\ee
This operator is conjectured to have the open topological string wave function on $\cy$ in the NS limit, $\ZtopopenNS$, in its kernel,
\be \label{kernel}
\cO_{\mirrorcurve} \,\ZtopopenNS = 0 \,,
\ee
with $x$ identified as the open string modulus. \cite{Franco:2015rnr} identifies the equation 
\be
\cO_{\mirrorcurve} \Psi = 0  \label{kernel_psi}
\ee
as the quantum Baxter equation for the Goncharov-Kenyon integrable system determined by the toric data of $\cy$. The eigenvalues of the Hamiltonians of this system map to the complex structure parameters $z_i$, $i=1, \ldots, g$, of $\mirrorcurve$. Solving the quantum Baxter equation (\ref{kernel_psi}) with appropriate boundary conditions on $\Psi$ is equivalent to solving the spectral problem. Numerical evidence for this beyond the genus one case is reported in \cite{Hatsuda:2015qzx}.

The conjectured quantization condition \cite{Franco:2015rnr} is a set of equations, indexed by $g$ integers $n_i$, whose solution set of $g$-tuples is to coincide with the Goncharov-Kenyon spectrum. The ingredients that enter into the quantization condition are the Nekrasov-Shatashvili limit of the refined topological string free energy \cite{Iqbal:2007ii,Nekrasov:2009rc}, $\FNS$, and the quantum mirror map \cite{Aganagic:2011mi}. $\FNS$ encodes integer invariants associated to the Calabi-Yau $\cy$ \cite{Choi:2012jz}. These appear most naturally when it is expressed in terms of the flat coordinates $T_i$ on the complexified K\"ahler moduli space of $\cy$. We can distinguish between two contributions to $\FNS$. First, there is a perturbative contribution which depends on the triple intersection numbers $a_{ijk}$ of the compact toric divisors of $\cy$ (suitably generalized to the non-compact setting) and integers $b_i^{NS}$, which have not been given a geometric interpretation yet,
\be
\FNSpert({\boldsymbol{T}},\hbar) = \frac{1}{6 \hbar} \sum_{i,j,k=1}^{g_N} a_{ijk} T_i T_j T_k + \left( \frac{4 \pi^2}{\hbar} \right) \sum_{i=1}^{g_N} b_i^{NS} T_i  \,.
\ee
The second contribution depends on integer invariants $N_{j_L,j_R}^{\boldsymbol{d}}$ of the geometry, with $\boldsymbol{d}$ a $g_N$-tuple mapping to a class in $H_2(\cy)$ via the choice of coordinates $T_i$, and the half-integers $(j_L,j_R)$ indicating a representation of $SU(2) \times SU(2)$, and has the form
\be \label{F_ns}
\FNSBPS({\boldsymbol{T}}, \hbar) = \sum_{w=1}^\infty \sum_{j_L,j_R} \sum_{{\boldsymbol{d}}} \frac{N_{j_L, j_R}^{\boldsymbol{d}}}{2 w^2} \frac{\sin[\frac{\hbar w}{2}(2j_l+1)] \sin[\frac{\hbar w}{2}(2j_R+1)]}{\sin^3 \frac{\hbar w}{2}} \boldsymbol{Q}^{w {\boldsymbol{d}}} \,,
\ee
with $Q_i = \exp(-T_i)$ as introduced above, and $\boldsymbol{Q}=(Q_1,\ldots,Q_g)$. Following \cite{Franco:2015rnr}, we have indexed this contribution with ${}^{\mbox{\tiny BPS}}$ due to its enumerative interpretation \cite{Gopakumar:1998jq,Iqbal:2007ii,Huang:2010kf,Choi:2012jz}. In the spirit of \cite{Nekrasov:2009rc}, one would then like to impose a quantization condition on the parameters $Q_i$ via
\be \label{qc_naive}
\sum_{j=1}^{g_N} C_{ij} \frac{\partial}{\partial T_j} \Big( \FNSpert({\boldsymbol{T}},\hbar) + \FNSBPS({\boldsymbol{T}},\hbar) \Big) = 2 \pi \left(n_i + \frac{1}{2} \right) \,.
\ee
$C_{ij}$ is the intersection matrix between a basis of curve classes in $\cy$, corresponding to a basis of the Mori cone of the toric geometry and the coordinates $T_i$, and the torically invariant divisors of $\cy$. It arises in \cite{Hosono:2004jp} to relate the derivatives of the prepotential to these divisors.

The crucial ingredient in the quantization condition of \cite{Franco:2015rnr}, inspired by the so-called pole cancellation mechanism in \cite{Hatsuda:2012dt}, is to consider a third contribution to the quantization condition based on $\FNSBPS$, but evaluated at (up to a detail to which we return presently) 
\be \label{pole_cancellation}
(\frac{2\pi}{\hbar}{\boldsymbol{T}}, \frac{4 \pi^2}{\hbar}) \,. 
\ee 
The inspiration behind including this term stems from the observation that the contribution (\ref{F_ns}) to the free energy has poles, due to the sum over $w$, at $\hbar= 2\pi \frac{r}{s}$ for all integer values of $r$ and $s$. As a function of $q=\exp(i\hbar)$, it hence necessarily exhibits at best a natural boundary of analyticity on the unit circle (in fact, we will see in the example of local $\IP^1 \times \IP^1$ in section \ref{subsection:numerics} that even away from the unit circle, the expansion (\ref{F_ns}) is not convergent). The quantization condition (\ref{qc_naive}) as it stands is hence ill-defined, at least for such values of $\hbar$. The evaluation point (\ref{pole_cancellation}) is chosen to precisely cancel the contribution from each of these poles: for $\hbar= 2\pi \frac{r}{s}$, the pole which arises in (\ref{qc_naive}) at $w= ls$, $l \in \IN$ is canceled by the contribution of the corresponding derivative of the NS free energy evaluated at (\ref{pole_cancellation}) at $w= lr$, $l \in \IN$. This almost works as is: the residues evaluate to
\begin{align} \label{residue_pert}
-\Res&|_{\hbar= 2\pi \frac{r}{s}} \frac{d_j N_{j_L, j_R}^{\boldsymbol{d}}}{2 w} \frac{\sin[\frac{\hbar w}{2}(2j_L+1)] \sin[\frac{\hbar w}{2}(2j_R+1)]}{\sin^3 \frac{\hbar w}{2}} \boldsymbol{Q}^{{\boldsymbol{d}}\,w}|_{w=ls} = \\ 
&=-\frac{d_j N_{j_L, j_R}^{\boldsymbol{d}}}{w^2} (2j_L+1)(2j_R+1) \frac{\cos[\frac{\hbar w}{2}(2j_L+1)] \cos[\frac{\hbar w}{2}(2j_R+1)]}{\cos^3 \frac{\hbar w}{2}} \boldsymbol{Q}^{{\boldsymbol{d}}\,w} |_{\hbar= 2\pi \frac{r}{s},w=ls} \nn\\
&=-\frac{d_j N_{j_L, j_R}^{\boldsymbol{d}}}{(ls)^2} (2j_L+1)(2j_R+1) (-1)^{lr(2j_L+2j_R+1)} \boldsymbol{Q}^{{\boldsymbol{d}}\,ls } \nn 
\end{align}
and
\begin{align} \label{residue_np}
- \Res|_{\hbar= 2\pi \frac{r}{s}}\frac{d_j N_{j_L, j_R}^{\boldsymbol{d}}}{2 w} &\frac{\sin[\frac{2\pi^2 w}{\hbar}(2j_l+1)] \sin[\frac{2\pi^2 w}{\hbar}(2j_R+1)]}{\sin^3 \frac{2\pi^2 w}{\hbar}} \boldsymbol{Q}^{{\boldsymbol{d}}\,\frac{2 \pi}{\hbar} w }|_{w=lr} = \\ 
&=\frac{d_j N_{j_L, j_R}^{\boldsymbol{d}}}{(ls)^2} (2j_L+1)(2j_R+1) (-1)^{ls(2j_L+2j_R+1)} \boldsymbol{Q}^{{\boldsymbol{d}}\,ls} \,. \nn
\end{align}
The sign factors in (\ref{residue_pert}) and (\ref{residue_np}) can be adjusted such that the two terms cancel if the K\"ahler parameters can be shifted by a $B$-field that satisfies
\be \label{B_shift_closed}
(-1)^{2j_L+2j_R +1 + \boldsymbol{B}\cdot{\boldsymbol{d}}} = 1 
\ee 
for all pairs $(j_L,j_R)$ for which $N_{j_L,j_R}^{\boldsymbol{d}} \neq 0$ \cite{Hatsuda:2013oxa}. The existence of such a $B$-field has been shown for many classes of examples, but a proof of its existence for all toric geometries is still lacking. Combining these elements yields the conjectured quantization condition
\begin{multline} \label{quantization_condition} 
\sum_{j=1}^{g_N} C_{ij} \frac{\partial}{\partial T_j} \Big[ F_{NS,pert}({\boldsymbol{T}},\hbar) + F_{NS,BPS}({\boldsymbol{T}}+ \pi i \boldsymbol{B},\hbar) + \\ +\frac{\hbar}{2\pi}F_{NS,BPS}\Big(\frac{2 \pi}{\hbar}{\boldsymbol{T}}+ \pi i \boldsymbol{B},\frac{4 \pi^2}{\hbar}\Big)\Big] = 2 \pi \left(n_i + \frac{1}{2} \right) \,.
\end{multline}
The equations (\ref{quantization_condition}) can be solved to express the K\"ahler parameters ${\boldsymbol{T}}$ in terms of the integers $n_i$. The so-called quantum mirror map, discussed further in section \ref{section:example}, then maps these solutions to the eigenvalues $z_i$ of the Goncharov-Kenyon spectral problem.\footnote{A subtle shift in the mirror map is required at non-vanishing $B$-field \cite{Franco:2015rnr}. In the case of local $\IP^1 \times \IP^1$ that we consider in section \ref{section:example}, $\boldsymbol{B}=0$, hence this shift does not arise.\label{footnote:shift}}

\subsection{The open topological string partition function} \label{subsection: Z_open}
The open topological string partition function, as defined in \cite{Witten:1992fb}, serves as a generating function for open Gromov-Witten invariants, counting maps, in an appropriate sense, from Riemann surfaces with boundary to a Calabi-Yau manifold $\cy$ with branes on which these boundaries are constrained to lie. We will call this partition function $\ZtopopenGW = \exp \FtopopenGW$. When $\cy$ is a toric Calabi-Yau manifold, this notion can be refined \cite{Gukov:2004hz,Aganagic:2011sg}, and leads to a formal series in two expansion parameters $\epsilon_1$ and $\epsilon_2$.

Beginning with \cite{Aganagic:2003qj}, it has been gradually understood  \cite{Mironov:2009uv,Aganagic:2011mi} that the monodromy of the open topological string partition function is intimately related to the corresponding closed topological string partition function. The mirror curve $\mirrorcurve$ to the toric Calabi-Yau $\cy$ is identified as the open string moduli space \cite{Aganagic:2000gs,Aganagic:2001nx}, such that $\FtopopenGW$ becomes a function on $\mirrorcurve$. The leading order contribution to $\FtopopenGW$ in an $\epsilon_{1,2}$ expansion is then given by
\be
\FtopopenGW \sim \int^x \lambda \,,
\ee
where $\lambda$ is the meromorphic 1-form introduced in section \ref{subsection:qc}. Thus, the monodromy of this leading contribution around the $A$- and $B$-cycles of the mirror curve coincide with the periods of $\lambda$. These determine the prepotential $F_0$ of $\cy$ via the special geometry relations
\be \label{special_geometry}
\left.
\begin{array}{ll}
	T_i(\bz) &= \oint_{A_i} \lambda \\
	T_{D,i}(\bz) &= \oint_{B_i} \lambda
\end{array}
\right\}
\quad T_{D,i}(\bz(\boldsymbol{T})) = \frac{\partial}{\partial{T_i}} F_0(\boldsymbol{T}) \,.
\ee
In refined topological string theory, $F_0$ is the leading contribution in the formal expansion of $\Ftop$ in $\epsilon_1$ and $\epsilon_2$. It was argued in \cite{Aganagic:2011mi}, based on insights from \cite{Aganagic:2003qj,Nekrasov:2009rc,Mironov:2009uv}, that the higher order corrections to $\Ftop$ in the NS limit\footnote{This corresponds to the expansion of $\Ftop$ in $s = (\epsilon_1 + \epsilon_2)^2$ at leading order in $g_s^2 = \epsilon_1 \epsilon_2$. In this limit, it has become conventional to denote $\epsilon_1 = \hbar$, as we have done in section \ref{subsection:qc}.} $\epsilon_2 \rightarrow 0$, should arise as the monodromy of $\FtopopenGWNS$, given by 
\be \label{fopenlambda}
\FtopopenGWNS \sim \int^x \lambdaq \,,
\ee
with $\lambdaq$ identified with the exponent $S$ of the WKB ansatz discussed in section \ref{section:monodromy_from_WKB}.
The special geometry relation (\ref{special_geometry}) now takes the form
\be \label{NS_conjecture}
\left.
\begin{array}{ll}
	T_i(\bz) &= \oint_{A_i} \lambdaq \\
	T_{D,i}(\bz) &= \oint_{B_i} \lambdaq
\end{array}
\right\}
\quad T_{D,i}(\bz(\boldsymbol{T})) = \frac{\partial}{\partial{T_i}} \Fns (\boldsymbol{T}) \,.
\ee
This proposal was checked explicitly in \cite{Mironov:2009uv} for pure 4d $SU(2)$ gauge theory. In the framework of the AGT correspondence \cite{Alday:2009aq}, the necessity to take the NS limit to relate $\Ftopopen$ to $\lambdaq$ becomes particularly transparent, see \cite{KashaniPoor:2012wb}. (\ref{NS_conjecture}) was further checked in both the 4d and 5d setting in \cite{Huang:2011qx,Huang:2012kn,Huang:2014nwa}. It was shown to follow from the AGT correspondence in \cite{Kashani-Poor:2014mua} for $\cN=2^*$ gauge theory.

The refined open topological string partition function on toric geometries can also be defined as an index. In this incarnation, it takes the form \cite{Gukov:2004hz,Aganagic:2011sg}
\be \label{F_top_open}
\FtopopenBPS(q,t,\mathbf{Q},x) = -\sum_{n=0}^\infty\sum_{s_1,s_2,{\boldsymbol{d}}} {\sum_{m \neq 0}}  D_{m,{\boldsymbol{d}}}^{s_1,s_2} \frac{q^{n s_1}t^{-n s_2}}{n(1-q^n)}\mathbf{Q}^{n{\boldsymbol{d}} }e^{m n \hat{x}} 
\ee
in an expansion in the appropriate exponentiated coordinates on the open and closed string moduli space. $\hat{x}$ denotes
the open string modulus encoding the position of the brane and the value of a $U(1)$ Wilson line along the boundary of the topological string worldsheet. $\mathbf{Q}=(Q_1, \ldots, Q_n)$ are the exponentials of the flat closed string moduli $\boldsymbol{t}$ appearing in (\ref{NS_conjecture}), and $q= e^{i \epsilon_1}$, $t= e^{- i \epsilon_2}$. The $\,\hat{}\,$ over the open string modulus is to indicate that a naive choice of this coordinate must be modified by factors of closed string moduli in order to obtain integer open string invariants $D_{m,{\boldsymbol{d}}}^{s_1,s_2}$. The need for such so-called flat open coordinates was first exposed in \cite{Aganagic:2000gs,Aganagic:2001nx} (see also \cite{Iqbal:2001kk}, where an alternate algorithm was proposed to compute these coordinates). In the Nekrasov-Shatashvili limit $t \rightarrow 1$, we set
\be
D_{m,{\boldsymbol{d}}}^{s_1} = \sum_{s_2} D_{m,{\boldsymbol{d}}}^{s_1,s_2} \,.
\ee

The expansion of $\FtopopenGW$ in open and closed string moduli coincides with that of $\FtopopenBPS$ in $\epsilon_1$ and $\epsilon_2$, giving rise to the conjecture that an underlying function $\Ftopopen$ should exist from which both descend. We hence ask to what extent the relation between the monodromy of the open topological string and $\Ftop$ persists in the expansion $\FtopopenBPS$.

For simplicity, let us restrict the discussion to the case of genus 1 mirror curves $\mirrorcurve$. The $A$-cycle in these geometries is given by the phase of the open string modulus $X= \exp(x)$. The $B$-cycle is encoded in the branch cut structure of $\lambda$ as a function of $X$. $\FtopopenBPS$ clearly does not contribute to the $A$-monodromy of $\Ftopopen$ under $x \rightarrow x + 2 \pi i$. Indeed, this monodromy is due to a contribution $\FtopopenLin$ to $\Ftopopen$ which is linear in $x$, and which can easily be computed from the quantum curve $\cO_{\boldsymbol{C}}$ \cite{Aganagic:2011mi,Huang:2014nwa}. Thus, in an $X$ expansion, 
\be \label{X_exp}
\Ftopopen \sim_X \FtopopenLin + \FtopopenBPS \,,
\ee
with the $A$-monodromy due exclusively to the first term on the RHS. As the $B$-monodromy is due to the branch cut structure in $X$, it is not visible upon expanding in $X$. This can be seen at leading order in $\epsilon$ by studying $\int^x \lambda$. Hence, the $B$-monodromy should be determined after combining the two terms in (\ref{X_exp}) by first summing the infinite series in $\exp(x)$ of their $x$ derivative.

\subsubsection{The dependence on $\hbar$} \label{section:fcn_eps}
The convergence of the sums in (\ref{F_top_open}) over ${\boldsymbol{d}}$ and $m$ depends on the growth properties of the constants $D_{m,{\boldsymbol{d}}}^{s_1,s_2}$. But already the sum over multi-wrappings $n$ is problematic: for $\frac{\hbar}{2\pi}  \in \IQ$, the summand diverges for infinitely many $n$. $\FtopopenBPS$ as presented in (\ref{F_top_open}) hence exhibits poles at a dense set of points on the unit circle in the $q$-plane. This is the open string analogue of the behavior of the closed topological string amplitude discussed in section \ref{subsection:qc}. To study this phenomenon, we will begin by considering the quantum dilogarithm \cite{Faddeev:1993rs}. For $|X|<1$, this function can be defined via the exponential of an infinite sum,
\be
(X;q)_{\infty} = \exp \left[ -\sum_{k=1}^\infty \frac{X^k}{k} \frac{1}{1-q^k} \right]\,,
\ee
which takes the product form
\be
(X;q)_{\infty} = \begin{cases}
	\prod_{n=0}^\infty (1- Xq^n)  \quad \mathrm{if} \,\,\, |q|<1 \,,\\
	\prod_{n=0}^\infty \frac{1}{ (1-X q^{-(n+1)})} \quad \mathrm{if} \,\,\, |q|>1 \,. 
\end{cases}
\ee
$(X;q)$ converges uniformly inside and outside the unit circle on the $q$-plane, but is ill-defined on a dense subset of the unit circle itself. To address this problem, Faddeev introduced what he called the modular quantum dilogarithm in \cite{Faddeev:1995nb}, by considering the quotient (in our notation)
\be \label{modular_q_dilog}
\gamma(x,\hbar)=\frac{(X;q)_{\infty}}{(X^{\frac{2\pi}{\hbar}};q^{-\frac{4\pi^2}{\hbar^2}})_{\infty}} = \frac{(e^{x};e^{i \hbar})_{\infty}}{(e^{\frac{2\pi x}{\hbar}};e^{-\frac{4 \pi^2 i}{\hbar}})_{\infty}}\,.
\ee
The denominator is chosen to cancel the poles of the numerator. To see the mechanism at work, consider such a pole at $ \frac{\hbar}{2\pi} = \frac{r}{s}$. The sum entering in the dilogarithm in the numerator has a summand at $k=m s$ that exhibits a pole with residue
\be
-\frac{X^k}{k} \frac{1}{1-q^k} \sim -\frac{X^{ms}}{ms} \frac{1}{-  i ms(\hbar - 2\pi \frac{r}{s})}  \,.
\ee
A corresponding term in the denominator which cancels this contribution stems from the summand at $k= mr$, with residue
\be
-\frac{X^{\frac{2\pi k}{\hbar}}}{k} \frac{1}{1-q^{-\frac{4 \pi^2 k}{\hbar^2}}} \sim -\frac{X^{ms}}{mr} \frac{1}{ -\frac{ imr}{(\frac{r}{s})^2} (\hbar - 2\pi\frac{r}{s})}  \,.
\ee
By re-ordering the two formal infinite sums that occur in the exponentials of (\ref{modular_q_dilog}), we obtain a function defined everywhere on the $q$-plane which coincides with the product (\ref{modular_q_dilog}) for $q$ off the unit circle.

Note that the pole cancellation mechanism works for any sum of the form 
\be \label{gen_open_pol_can}
\sum_k \frac{f_k(q^k,t^k, \boldsymbol{Q}^k,X^k)}{k(1-q^k)} \,,
\ee
for $f_k$ a rational function of its arguments, by subtracting a contribution
\be \label{gen_open_pol_can_np}
\sum_k \frac{f_k(q^{-\frac{4\pi k}{\hbar^2}},t^{\frac{2\pi k}{\hbar}}, \boldsymbol{Q}^{\frac{2\pi k}{\hbar}},X^{\frac{2\pi k}{\hbar}})}{k(1-q^{-\frac{4\pi k}{\hbar^2}})} \,,
\ee
i.e. as long as all parameters aside from $q$ in the correction term are evaluated to the power of $\frac{2\pi}{\hbar}$. Returning to the $\exp(x)$ expansion of the open topological string partition function (\ref{F_top_open}), we note that $\ZtopopenBPS$ is almost of the form (\ref{gen_open_pol_can}), up to the fact that $q$ and $t$ are evaluated to half-integer powers. Running through the pole cancellation argument for this case, we see that half-integer powers of $q$ lead to a sign factor at $\hbar = 2\pi \frac{r}{s}$, $k=ms$,
\be
q^{k s_1} = (-1)^{2 s_1 m r} \,,
\ee
and likewise in the correction term,
\be
q^{-\frac{4 \pi^2}{\hbar^2} k s_1} = (-1)^{2 s_1 m s} 
\ee
at $k=mr$. For the cancellation mechanism to work, we can shift the K\"ahler parameters by a B-field that satisfies
\be \label{B_shift_open}
(-1)^{2s_1 + \boldsymbol{B \cdot d}}=1
\ee
for all $s_1$ for which $D_{m,\bd}^{s_1,s_2} \neq 0$.

\subsubsection{Specifying the domain $\functionspace$ of $\cO_{\mirrorcurve}$} \label{subsection:domain}
As reviewed in section \ref{subsection:qc}, $\ZtopopenNS$ is conjectured to be annihilated by the operator $\cO_{\mirrorcurve}$.

The simplest instance of this behavior can be observed for the open topological string partition function on $\IC^3$, which is given by the quantum dilogarithm introduced above. It satisfies the difference equation
\be \label{quantum_curve_c3}
\left[(1-e^x) -  e^{p}\right] (e^x;e^{i \hbar})_{\infty} =0\,.
\ee
Note that a difference equation of the form 
\be \label{diff_equ}
\cO_{\mirrorcurve} \Psi(x) = 0
\ee
does not have a unique solution. In particular, given a function $\Psi(x)$ in the kernel of the operator $\cO_{\mirrorcurve}$, $\chi(x) \times \Psi(x)$ for any function $\chi(x)$ of periodicity $ i \hbar$ will also be annihilated by this operator. In the case of $\cy=\IC^3$, the modular quantum dilogarithm (\ref{modular_q_dilog}) is hence also annihilated by the operator on the LHS of (\ref{quantum_curve_c3}).

We know three methods to determine an element in the kernel of $\cO_{\mirrorcurve}$. The first proceeds via the WKB ansatz
\be \label{wkb_ansatz}
\psiw(x) = \exp  \int^x S  \,, \quad  S= \frac{1}{\epsilon}S_{-1} + \sum_{n=0}^\infty S_n \epsilon^n 
\ee
we discussed in section \ref{section:monodromy_from_WKB}, with $\epsilon= i \hbar$. Substituting (\ref{wkb_ansatz}) into (\ref{diff_equ}) allows us to recursively solve for the coefficients $S_n$. The formal power series thus obtained, which we have called $\psiw$, should coincide, up to the ambiguity discussed above, with $\FtopopenGWNS$. The second method is based on rewriting (\ref{diff_equ}) as an equation for
\be \label{quotient_for_rec}
\V(x) = \frac{\Psi(x- i\hbar)}{\Psi(x)} \,,
\ee
which can then be solved recursively and yields an expansion of $\V(x)$ in the closed moduli parameters $z_i$. We can extract $\Psi(x)$ from $\V(x)$ up to the ambiguity discussed. Expressing the moduli $z_i$ in $\V(x)$ in terms of flat coordinates $T_i$, this should coincide with $\ZtopopenBPSNS$. We therefore refer to this formal series in $z_i$ (upon a choice of the ambiguity) as $\psiBPS$. We will apply both methods to the example of local $\IP^1 \times \IP^1$ in section \ref{section:example}.

The ambiguity of multiplying $\Psi(x)$ via a periodic function in $x$ can be reduced by specifying the function space on which the operator $\cO_{\mirrorcurve}$ acts. Aside from the behavior in $x$, the dependence on the parameters $q$ and $\bz$ needs to be specified.
The third method of computing $\Psi(x)$ explicitly depends on this choice of function space. It proceeds by specifying a basis for this space, expressing $\cO_{\mirrorcurve}$ as a matrix $\cO_{\mirrorcurve}^N$ in a truncation of this basis to $N$ elements. The values of $\bz$ for which the kernel of $\cO_{\mirrorcurve}^N$ is non-empty can then be determined by solving $\det \cO_{\mirrorcurve}^N = 0$, upon which the kernel in the approximation of this truncation easily follows.

The choice of function space made in \cite{Huang:2014eha,Hatsuda:2015qzx,Franco:2015rnr} is the $L^2(\IR)$ space spanned by the eigenstates of the harmonic oscillator,
\be \label{basis_harm_osc}
\psi_n(x) = \frac{1}{\sqrt{2^n n!}} \left(\frac{m \omega}{\pi \hbar} \right)^{\frac{1}{4}} e^{- \frac{ m \omega x^2}{2 \hbar}} H_n( \sqrt{\frac{m \omega}{\hbar}} x) \,, \quad n\in \IN_{0} \,.
\ee
The $H_n(x)$ are the Hermite polynomials, and $\omega$ and $m$ are physical parameters which will play no role for our purposes and will be set to convenient values in the following. Upon computing the matrix elements of the monomials generated by $e^x$ and $e^p=e^{-i \hbar \frac{\partial}{\partial x}}$,
\be  \label{harmonic_expectation}
\langle k | e^{ax} e^{bp} | l \rangle = 2^{\frac{k+l}{2}} \sqrt{k! l!}e^{|\zeta|^2 + i \frac{ab \hbar}{2}} \zeta^k \bar{\zeta}^l \sum_{n=0}^{\min\{k,l\}} \frac{2^n}{n!(k-n)!(l-n)!}\frac{1}{(2|\zeta|)^{2n}} \,,
\ee	
where
\be
\zeta= \frac{1}{2} \left(\frac{\hbar}{m \omega}\right)^{\frac{1}{2}} \left(a + i m \omega b \right) \,,
\ee
the matrix elements of operators $\cO_{\mirrorcurve}$ in this basis can easily be determined. By making the choice of basis (\ref{basis_harm_osc}), we are committing to a certain type of $\hbar$ dependence. The states (\ref{basis_harm_osc}) depend continuously on $\hbar$ (up to branch cuts) and are defined for any value $\hbar \in \IC^*$. They are elements of $L^2(\IR_x)$ for $\re(\hbar) > 0$. The kernel of $\cO_{\mirrorcurve}^N$ is determined by solving a system of $N$ linear equations with coefficients the matrix elements (\ref{harmonic_expectation}). The solution will be a linear combinations of the harmonic oscillator eigenstates (\ref{basis_harm_osc}) with coefficients that are rational functions of these matrix elements. We will call $\functionspace$ the space of functions of the variables $(\hbar,x)$ of this form. The quantization condition presented in section \ref{subsection:qc} is to yield the tuples $\bz$ for which the kernel of the operator $\cO_{\mirrorcurve}$ has non-zero intersection with this function space.


\subsubsection{Consequences of imposing $\Psi \in \functionspace$} \label{subsection: consequences qc}
Let us assume that $\psiw$, the formal power series in $\epsilon= i\hbar$ defined in (\ref{wkb_ansatz}), can be Borel resummed to a function $\Psiw$ away from $q=1$. For $\Psiw$ to be an element of $\functionspace$, it must be single-valued as a function of $x$. We have argued in section \ref{section:monodromy_from_WKB} that the monodromy along a path $\cC$ is given by $\exp \Pi_{\cC}$, with $\Pi_{\cC}$ the Borel resummation of the integral $\pi_{\cC} = \oint_{\cC} S$, and $S$ defined in (\ref{wkb_ansatz}). When $\cC$ coincides with the $B_i$-cycle of the geometry, the NS conjecture (\ref{NS_conjecture}) identifies $\pi_{\cC}$ with $\partial_{t_i} \Fns$. The monodromy is thus of the form $\exp[\phipert(\bz,q)+\phiBPS(\bz,q)]$. The condition for single-valuedness of $\Psiw$ around the cycle $B_i$ is hence
\be
\phipert(\bz,q)+\phiBPS(\bz,q) = 2 \pi i \, n \,, \quad n\in \IZ \,.
\ee
In a $z_i$ expansion, $\psiw$ reproduces the $\hbar$ expansion of $\psiBPS$, which was defined below (\ref{quotient_for_rec}). While for real $\frac{\hbar}{2\pi} \notin \IQ$, it has been noted \cite{Grassi:2014cla,Hatsuda:2015owa} that the Borel resummation of $\psiw$ is locally smooth in $\hbar$, there exists no argument that Borel resummation at complex $\hbar$ will eliminate the poles in $q$ plaguing $\psiBPS$. If this indeed does not occur, $\Psiw \notin \functionspace$. 

To proceed, we will assume that $\psiBPS$ is also Borel summable, in its expansion parameters $\bz$, to the function $\PsiBPS$. To enforce smooth behavior upon approaching the unit $q$-circle, we take our cue from the discussion in \ref{section:fcn_eps} and consider the quotient
\be \label{final_honcho}
\Psi = \frac{\Psiw(x)}{\PsiBPS(X^{\frac{2\pi}{\hbar}},\bz(\bQ^\frac{2\pi}{\hbar}), q^{-\frac{4\pi^2}{\hbar^2}})} \,.
\ee
Due to the periodicity of the denominator under $x \mapsto x + i \hbar$, this is still a solution to the difference equation (\ref{diff_equ}).\footnote{Note that the conditions of periodicity and pole cancellation do not fix the modification uniquely. Indeed, after the first version of this paper was submitted to the arXiv, we were informed of work in progress \cite{Marino:2016plus} suggesting a different completion.} The condition for the single-valuedness of $\Psi$ upon circumnavigating the cycle $B_i$ is given by
\be 
\phipert(\bz(\bQ),q)+\phiBPS(\bz(\bQ),q)  = 2\pi i m + \kappa \,,\quad \phiBPS(\bz(\bQ^{\frac{2\pi}{\hbar}}),q^{-\frac{4\pi^2}{\hbar^2}})= 2\pi i n+ \kappa \,,
\ee
for $m,n\in \IZ$, and arbitrary $\kappa \in \IC$, or equivalently,
\be 
\phipert(\bz(\bQ),q)+\phiBPS(\bz(\bQ),q) -\phiBPS(\bz(\bQ^{\frac{2\pi}{\hbar}}),q^{-\frac{4\pi}{\hbar^2}})  = 2\pi i (m-n)  \,.
\ee
We are here assuming that the perturbative contribution $\phipert(\bz(\bQ),q)$ to the quantization condition arises upon combining $\PsiBPS$ with an additional contribution, as in (\ref{X_exp}), and is not modified by the denominator of (\ref{final_honcho}). We will argue in section \ref{subsection:wkbp1p1} that the half-integer shift on the RHS of (\ref{quantization_condition}) is due to $\phipert$ containing a contribution $\phipert = \pi i + \ldots$. This explanation of the quantization condition predicts a relation between the closed invariants $N_{j_L, j_R}^{{\boldsymbol{d}}}$ and the open invariants $ D_{m,{\boldsymbol{d}}}^{s_1}$ based on the two conditions (\ref{B_shift_closed}) and (\ref{B_shift_open}).

\section{Example: local $\IP^1 \times \IP^1$} \label{section:example}
In this section, we will apply our analysis to the geometry $\cy=\cO(-K) \rightarrow \IP^1 \times \IP^1$. This geometry has been studied extensively in the literature with regard to its closed string invariants \cite{Katz:1996fh,Aganagic:2002wv,Choi:2012jz}, and in the context of the quantization condition (\ref{quantization_condition}) for real $\hbar$ \cite{Grassi:2014zfa}. Here, we will be interested in the WKB analysis of the difference equation (\ref{diff_equ}) for this geometry. As our analysis in section \ref{subsection: consequences qc} relies on complex $\hbar$, we will extend the study of (\ref{quantization_condition}) to this case.

In passing, we will also compute some open string invariants of this geometry and verify their integrality upon appropriate choice of flat open variables and invoking the quantum mirror map.

\subsection{The mirror curve and classical periods via Picard-Fuchs}
\begin{figure}
	\centering
	\includegraphics[width=5cm]{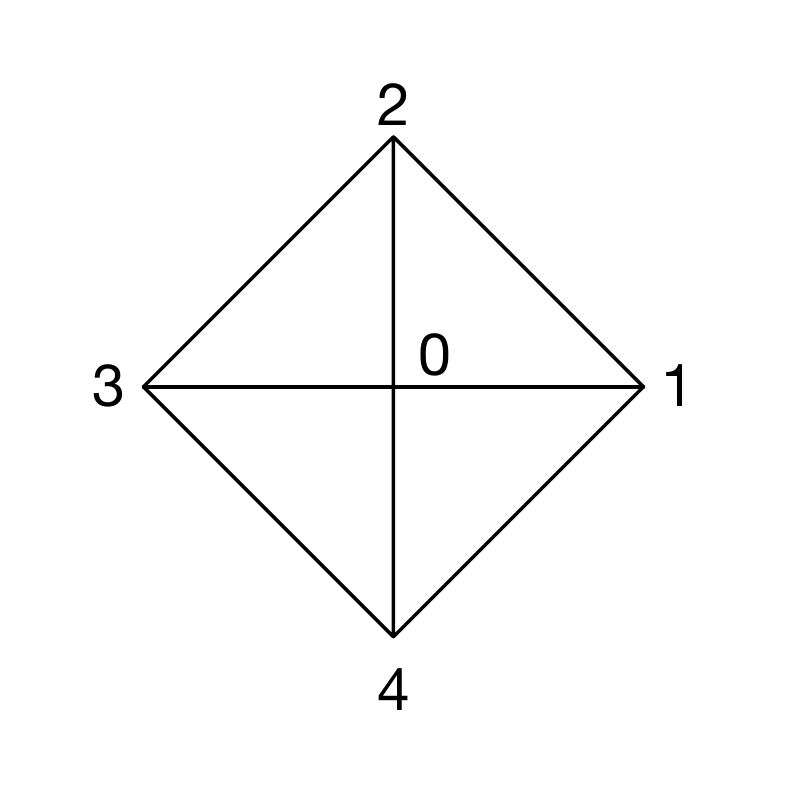}
	\caption{\small The toric grid diagram for local $\IP^1 \times \IP^1$.} \label{toric_diagram_p1xp1}
\end{figure}

The toric grid diagram describing the local $\IP^1 \times \IP^1$ geometry is depicted in figure \ref{toric_diagram_p1xp1},
with vertices corresponding to one dimensional cones of the fan enumerated from 0 to 4. The diagram exhibits one interior point and one boundary point beyond three. The underlying geometry is therefore described by one modulus and one mass parameter, in the terminology introduced in section \ref{subsection:qc}. Following the standard algorithm \cite{Chiang:1999tz}, each independent relation
\begin{align}
l^{(1)} &= ( -2 \,1\, 0\, 1\, 0 ) \\
l^{(2)} &= ( -2 \,0\, 1\, 0\, 1 ) 
\end{align}
among the one dimensional cones is assigned a parameter $z_i$, and the equation for the mirror curve $\mirrorcurve$ is obtained as
\be \label{mirrorcurvep1p1}
P_{\mirrorcurve}(e^x,e^p,e^{-x},e^{-p})=e^p + z_1 e^{-p} +  e^x + z_2 e^{-x} +1 = 0  \,.
\ee
Flat coordinates $t_1$ and $t_2$ on the complexified K\"ahler moduli space of $\cy$, encoding the size of the two $\IP^1$ curves respectively, are identified as the logarithmic solutions of the corresponding Picard-Fuchs system. These can be determined at small $z_1$, $z_2$ (the large radius regime on $\cy$) via the Frobenius method \cite{Chiang:1999tz} to be
\begin{eqnarray} 
-t_1 &=& \log z_1 + 2(z_1 + z_2) + 3(z_1^2 + 4 z_1 z_2 + z_2)^2 + \frac{20}{3}(z_1^3+9 z_1^2 z_2 +9 z_1 z_2^2 +z_2^3) +  \ldots \,, \nn \\
-t_2 &=& \log z_2 + 2(z_1 + z_2) + 3(z_1^2 + 4 z_1 z_2 + z_2)^2 + \frac{20}{3}(z_1^3+9 z_1^2 z_2 +9 z_1 z_2^2 +z_2^3) +  \ldots \,.\nn\\ \label{log_sol}
\end{eqnarray}
As above, we will also introduce exponentiated coordinates $Q_i = \exp (-t_i)$, such that small $z_i$ corresponds to large $t_i$ and small $Q_i$. The quotient $z_m = \frac{z_2}{z_1} = \frac{Q_2}{Q_1}$ is an algebraic function in the exponentials of the flat coordinates, identifying it as a mass parameter.

By inverting (\ref{log_sol}), we obtain the so-called mirror map
\be \label{mirror_map}
z_1 = Q_1 \Big( 1 - 2(Q_1 + Q_2) + (3Q_1^3 - 4 Q_1^2 Q_2 - 4 Q_1 Q_2^2 + 3Q_2^3 ) + \ldots \Big) \,. 
\ee
The doubly logarithmic solutions of the Picard-Fuchs system can also be determined via the Frobenius method at large radius, and allow the computation of the prepotential $F_0$ in this regime. Introducing coordinates $T = t_1$, $T_m = t_2-t_1$ to distinguish between modulus and mass parameter and expressing the doubly logarithmic solutions in terms of these, an appropriate linear combination of them yields $\partial_T F_0$. The correct linear combination can be determined e.g. by matching some low lying Gromov-Witten invariants (obtained e.g. by geometric means, or via the topological vertex).

\subsection{The quantum mirror curve and $\FtopopenNS$ via recursion} \label{fopen_p1xp1}
Following our discussion in section \ref{subsection:domain}, we introduce the function $\V(x) = \Psi(x - i \hbar)/\Psi(x)$ and rewrite the difference equation (\ref{diff_equ}) as an equation for $\V(x)$, 
\be \label{recursion}
\V(x) + \frac{z_1}{\V(x+ i \hbar)} + e^x + z_2 e^{-x} + 1 = 0 \,.
\ee
This equation can be solved recursively, yielding a formal series in $z_i$ which we call $\lV(x)$. Expressing $z_1$ and $z_2$ in terms of $Q_1$ and $Q_2$ via the quantum mirror map which we discuss below, we find
\be \label{three_cont}
\log \lV(x) = \vlin + \vBPSp + \vBPSn \,,
\ee
where 
\be
\vlin = Q_1  \,,
\ee
and $\vBPSp$, $\vBPSn$ are of the form\footnote{We thank Antonio Sciarappa for pointing out a sign error in formula (\ref{integerp1p1}) in a previous version of this paper.}
\be \label{integerp1p1}
\vBPSp(x,q) =  \sum_{n=0}^\infty \sum_{d_1,d_2,s_1} \sum_{m=1}^\infty D_{m,d_1,d_2}^{s_1} \frac{q^{n s_1}}{n}\frac{1-q^{-mn}}{1-q^n} Q_1^{n d_1} Q_2^{n d_2} e^{m n \hat{x}} \,,
\ee 
\be \label{BPS_symmetry}
\vBPSn(x,q) = \vBPSp(-x+\log Q_2,1/q) \,.
\ee
with integer coefficients $D_{m,d_1,d_2}^{s_1}$.\footnote{We have checked this structure up to $m=6$ and combined order 6 in $Q_1$ and $Q_2$.} $\hat{x}$ designates the flat open string modulus. It is given by \cite{Aganagic:2000gs,Iqbal:2001kk}
\be \label{open_flat}
e^{\hat{x}} = -\sqrt{\frac{Q_2}{z_2}} e^x \,.
\ee

Some invariants $D_{m,d_1,d_2}^{s_1}$ for low $m$ and $d_1+d_2$ are given in table \ref{table:invariants}. 

\begin{table} \label{table:invariants}
	\centering 
	\begin{tabular}{| c | c | c | c | c | }
		\hline
		$m$ & $d_2\, \setminus \,d_1$ & 0 & 1 & 2  \\
		\hline
		   & 0 &  & (2) 1 & (3) 1 \\
		1  & 1 &  & (2) 1 (3) 1& (1) 1 (2) 1 (3) 3 (4) 4 (5) 1   \\
		   & 2 &  & (2) 1 (3) 1 (4) 1 & (0) 2 (1) 4 (2) 4 (3) 8 (4) 11 (5) 11 (6) 4 (7) 1 \\
		\hline
		   & 0 &  & (3) 1 & (4) 1 (5) 1  \\
		2  & 1 & &  (3) 1, (4) 1& (2) 1 (3) 1 (4) 4 (5) 6 (6) 3 (7) 1    \\
		   & 2 & &  (3) 1 (4) 1 (5) 1& (1) 2 (2) 4 (3) 4 (4) 9 (5) 15 (6) 15 (7) 9 (8) 3 (9) 1  \\
		\hline
	\end{tabular}
	\caption{\small Some open string invariants $D_{m,d_1,d_2}^{s_1}$ for local $\IP^1 \times \IP^1$. The number in parentheses preceding the entry indicates the spin $s_1$.} \label{table:invariants}
\end{table}

As $D_{m,d_1,d_2}^{s_1} \neq 0$ only for integer $s_1$, we can choose $\boldsymbol{B}=0$ in (\ref{B_shift_open}). It is argued in \cite{Hatsuda:2015qzx} that this is also a valid choice in (\ref{B_shift_closed}).

\subsection{The quantum mirror curve and quantum periods}
The analytic structure of the mirror curve (\ref{mirrorcurvep1p1}) becomes clearer if we redefine variables by setting
\be \label{new_var}
\tilde{x} = x - \frac{1}{2} \log z_2 \,, \quad
\tilde{p} = p - \frac{1}{2} \log z_1 \,. 
\ee
This gives rise to the curve
\be \label{new_curve}
\cosh \tilde{p} + \sqrt{\frac{z_2}{z_1}} \cosh \tilde{x} + \frac{1}{2\sqrt{z_1}} = 0 \,.
\ee
Note that $\tilde{x}$ is essentially the flat coordinate on the open string moduli space give in (\ref{open_flat}). The question of the appropriate coordinates on this moduli space to achieve integrality of the expansion coefficients in (\ref{F_top_open}) is thus mapped to the question of the appropriate parametrization of the mirror curve. The relation (\ref{BPS_symmetry}) is a reflection of the symmetry of (\ref{new_curve}) under $\tx \leftrightarrow -\tx$.

\begin{figure}
	\centering
	\includegraphics[width=7cm]{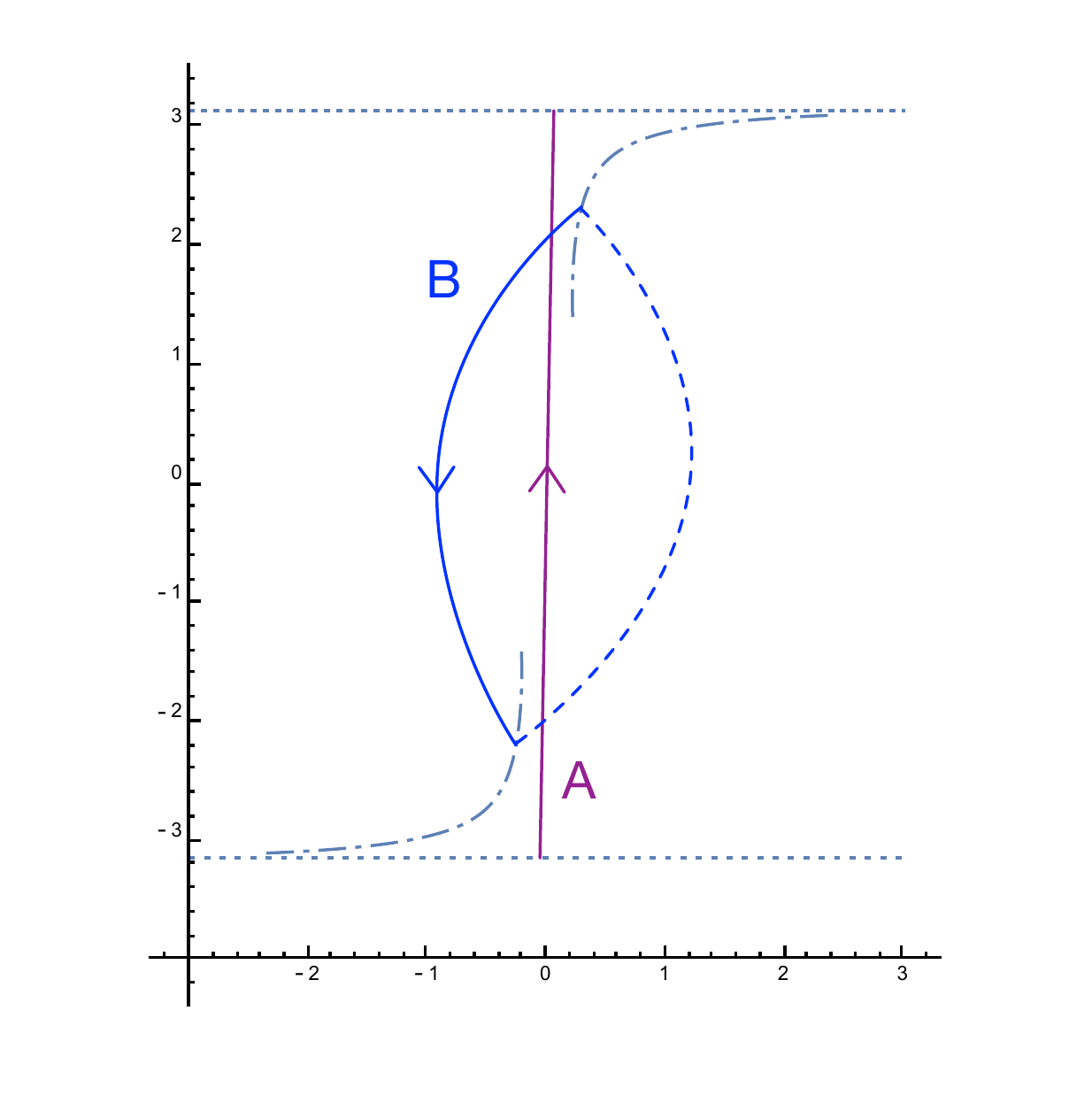}
	\caption{\small The sheet structure of the mirror curve for local $\IP^1 \times \IP^1$.} \label{mirror_p1xp1}
\end{figure}

Upon quantization, the shifts in (\ref{new_var}) preserve the canonical commutation conditions 
\be
[ x, p ] = i \hbar \quad \Rightarrow \quad [\tilde{x},\tilde{p}] = i \hbar \,.
\ee
The kernels of the quantization of the curves (\ref{mirrorcurvep1p1}) and (\ref{new_curve}) are related via
\ba \label{cosh_form}
&& \left[\cosh \tilde{p} + \sqrt{\frac{z_2}{z_1}} \cosh \tilde{x} + \frac{1}{2\sqrt{z_1}} \right] \tilde{\Psi}(\tilde{x})= 0  \\
& \Leftrightarrow& e^{\frac{-i x \log z_1}{2 \hbar}} \left[ e^p + z_1 e^{-p} +  e^x + z_2 e^{-x} +1\right] e^{\frac{i x \log z_1}{2 \hbar}} \tilde{\Psi}(x - \frac{1}{2} \log z_2)  = 0  \,.
\ea 
Equation (\ref{cosh_form}) expresses the quantum mirror curve in the appropriate form to map the equation (\ref{kernel_mirror_curve}) to a spectral problem of the form (\ref{spectral_problem_mirror_curve}), with
\be \label{spectralp1p1}
\tcO_{\mirrorcurve} = \cosh \tilde{p} + \sqrt{\frac{z_2}{z_1}} \cosh \tilde{x} \,.
\ee
As the leading contribution $S_{-1}$ to the WKB ansatz (\ref{wkb_ansatz}) coincides with the solution of (\ref{cosh_form}) for $\tp$, the analytic structure of this curve is captured by
\be
S_{-1}(\tx) = \pm \arccosh \left( \frac{1}{2\sqrt{z_1}} (1 + 2 \sqrt{z_2} \cosh \tilde{x}) \right) + \pi i\,.
\ee 
The dependence of $S_{-1}$ on $\tx$ is via $\cosh \tx$, a fundamental domain of the function hence lies between $\im \tx = -\pi$ and $\im \tx = \pi$. Within this interval, $S_{-1}$ requires two branch cuts, in accord with the discussion of the sheet structure of the $\arccosh$ function in section \ref{section:wkb_gen}. We have sketched this sheet structure in figure \ref{mirror_p1xp1}. 
Following the discussion of section \ref{section:wkb_gen}, both branch cuts are divided into two segments: the initial segment is the preimage of the interval $[-1,1]$ under the argument of the $\arccosh$. Crossing this branch cut changes the sign of the function; it is associated to a branch point of order 2. Crossing the branch cut beyond this point takes one to a sheet with imaginary part shifted by $2 \pi$; it is associated to a branch point of order infinity. We can define two conjugate cycles on this geometry, labeled by $A$ and $B$ in figure \ref{mirror_p1xp1}. The $A$-cycle reflects the periodicity of $\cosh \tx$. The $B$-cycle passes through the order 2 segment of the branch cuts.

The quantum mirror map \cite{Aganagic:2011mi} is obtained by defining the $A$-cycle integral of the WKB exponent $S(x)$ as a flat coordinate. By our discussion in section \ref{subsection: Z_open}, this coincides with the conventional mirror map to leading order in $\epsilon=i\hbar$. It is possible to compute this integral to all orders in $\hbar$ in a $z_i$ expansion by noting
\be
\log \lV(x) \sim_\epsilon \int^x \big( S(x'- i \hbar) - S(x') \big) dx' = -i\hbar S(x) + \sum_{n=1}^\infty S^{(n)}(x) \frac{(-i\hbar)^{n+1}}{(n+1)!} \,.\\
\ee
The integrand on the RHS is understood in an expansion in $z_i$. The integral of $\log \lV(x)$ along the $A$-cycle is easy to perform. Only $\vlin$ in (\ref{three_cont}) contributes, and yields \cite{Aganagic:2011mi,Huang:2014nwa}
\begin{align}
\Pi_A= -\frac{1}{2 \pi i} \int_{x_0- \pi i}^{x_0+\pi i} &  \vlin \,dx= \nn \\
&-\left(\frac{1}{2} \log z_1 +(z_1+z_2) + \frac{3 q z_1^2 + 2 z_1 z_2 + 8 q z_1 z_2 + 2 q^2 z_1 z_2 + 3 q z_2^2}{2q} + \ldots \right) \,.
\end{align}
Comparing to the result (\ref{log_sol}) obtained via the Picard-Fuchs equation at leading order in $\hbar$ allows us to fix the normalization for the quantum corrected period to be
\be
T_1 = -2 \Pi_A \,.
\ee
Inverting this relation yields the quantum mirror map, the first terms of which are
\be \label{quantum_mirror_map}
z_1 = Q_1 \left(1- 2(Q_1+Q_2) + 3 Q_1^2 - 2 \frac{(1-q)^2}{q}Q_1 Q_2 + 3Q_2^2 + \ldots  \right)\,.
\ee 
This expression is used to obtain the expansion of $\vBPSp$ in $Q_i$ in (\ref{integerp1p1}).

The integral along the $B$-cycle is more difficult to perform directly, as the branch cuts degenerate in the limit of vanishing $z_1$ and $z_2$. The non-logarithmic contributions to this period can be obtained from $\log \lV$ by performing the indefinite integral over $x$ order by order in a $z_i$ expansion, and extracting the finite contribution at $x \rightarrow - \infty$ \cite{Aganagic:2011mi,Huang:2014nwa}. A more elegant computation of the period is clearly desirable. Note that an expansion in $\exp(x)$, as has been performed to obtain the form (\ref{F_top_open}), does not commute with this integration.

\subsection{Exact WKB analysis} \label{subsection:wkbp1p1}
To perform a WKB analysis along the lines of section \ref{section:wkb_gen}, we consider the curve in the form (\ref{cosh_form}), allowing us to identify 
\be
- Q(\tx) =  \sqrt{\frac{z_2}{z_1}} \cosh \tilde{x} + \frac{1}{2\sqrt{z_1}} \,.
\ee
This yields the WKB expansion coefficients
\ba
S_{-1}(\tx) &=& \arccosh\left(\sqrt{\frac{z_2}{z_1}} \cosh \tilde{x} + \frac{1}{2\sqrt{z_1}}\right) + i \pi \,, \label{s-1_p1xp1} \\
S_0(\tx) &=& - \frac{1}{4} \frac{d}{dx} \log \left[ \left(\sqrt{\frac{z_2}{z_1}} \cosh \tilde{x} + \frac{1}{2\sqrt{z_1}}\right)^2 -1 \right] \,,\\
& \ldots& \,. \nn
\ea
Performing an expansion around vanishing $z_1$ and $z_2$ yields
\ba
S_{-1}(\tx) &=& \pi i - \frac{1}{2} \log z_1 + 2\sqrt{z_2} \cosh(\tx) - \left(z_1 + 2 z_2 \cosh^2(\tx) \right)  + \ldots \,,  \\
S_0(\tx) &=& -2 z_1 \sqrt{z_2} \sinh(\tx) + 6 z_1 z_2 \cosh(\tx) \sinh(\tx) \ldots \,,\\
& \ldots& \,. \nn
\ea
This coincides with the $\hbar$ expansion of the results obtained by solving (\ref{recursion}) via recursion.

\begin{figure}
	\centering
	\includegraphics[width=7cm]{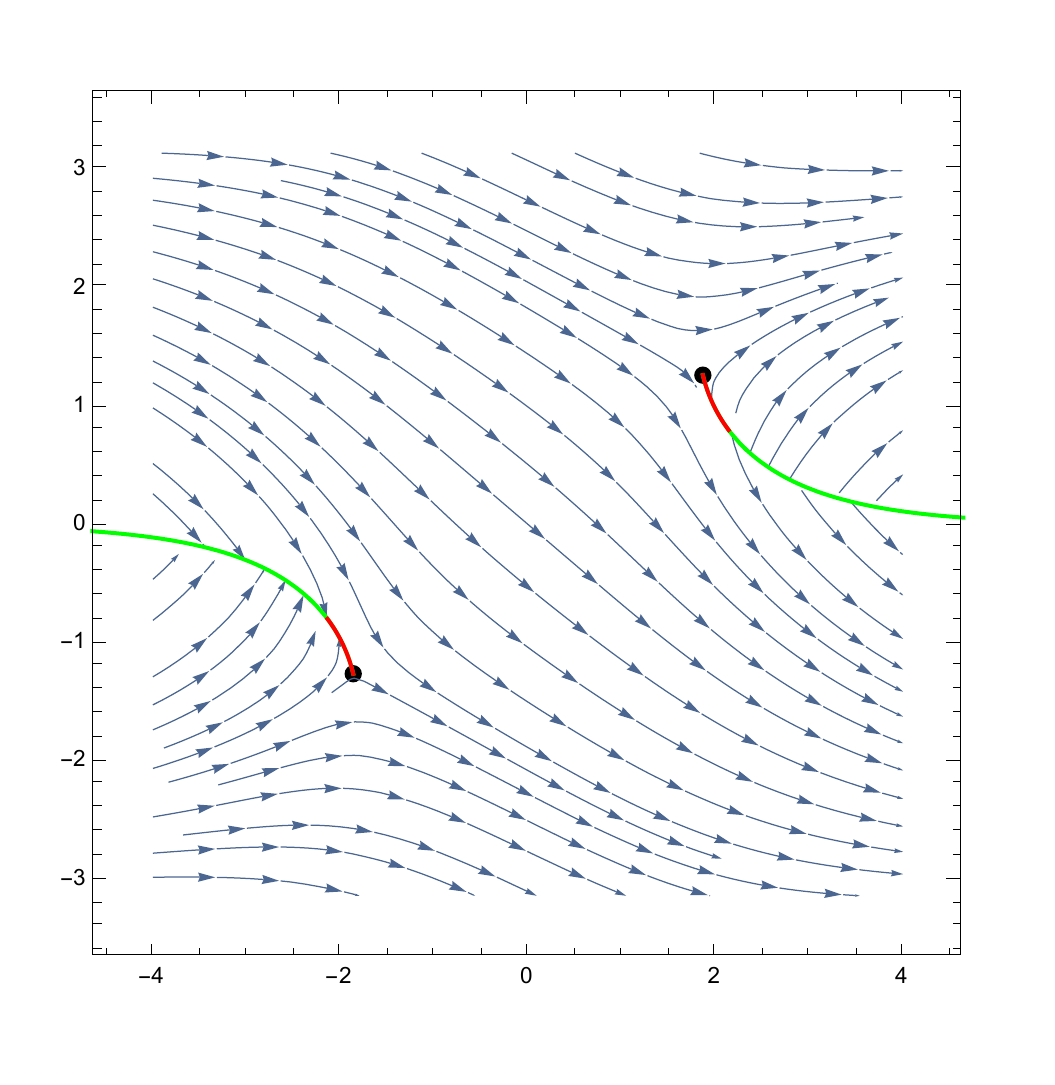}
	\caption{\small Flow lines for $S_{-1} = \arccosh(z-\cosh x)$. The branch cuts are drawn in red and green. The red sections are of order 2, the green sections of order $\infty$.} \label{figure_flow}
\end{figure}

\begin{figure}
	\centering
	\includegraphics[width=7cm]{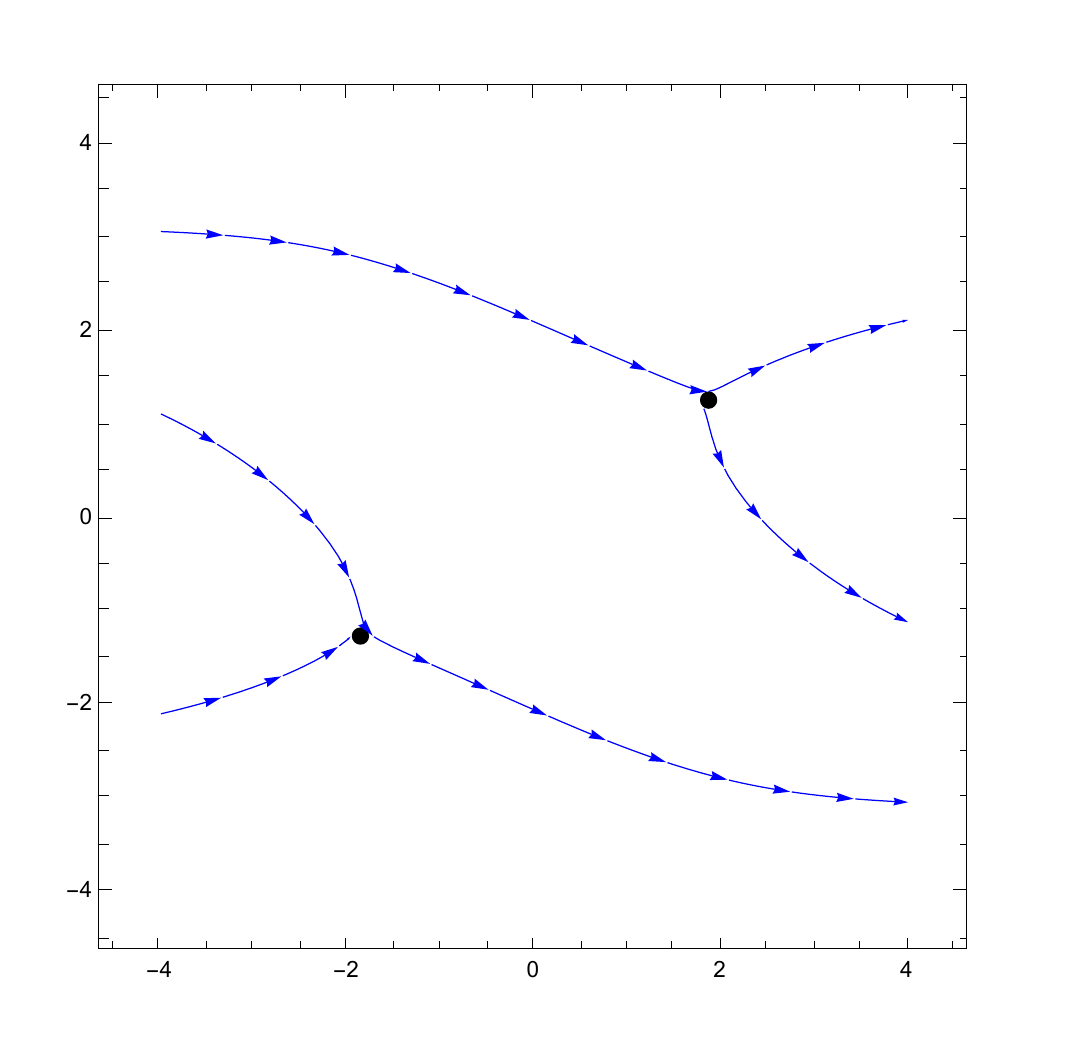}
	\caption{\small The Stokes graphs for local $\IP^1 \times \IP^1$.} \label{figure_SU2}
\end{figure}

The Stokes graphs of the difference equation are determined by (\ref{s-1_p1xp1}). There are two turning points in the range $\im x \in \{-\pi,\pi\}$, at $\sqrt{\frac{z_2}{z_1}} \cosh \tilde{x} + \frac{1}{2\sqrt{z_1}}= 1$. 
Figure \ref{figure_flow} shows the flow lines for $S_{-1}$, together with convenient choices for the branch cuts emanating from the turning points.
The curves $c(t)$ passing through the turning points for which
\be
e^{i\theta} S_{-1}dx \cdot \partial_t \in \IR \,,
\ee
for a given choice of $\theta$ and $z_1$, $z_2$, are depicted in figure \ref{figure_SU2}.
By our analysis of the previous section, these are the Stokes lines governing the monodromy behavior of $\Psiw^{\pm}$, at least close to the turning points: they pick up the monodromy $- \exp[\pm\oint S_{odd} dx]$ along a path circling the two turning points, with the integration performed along this path.
 
Note that the sign in front of the exponential in the monodromy is due to the fourth root arising from the $S_0$ contribution in (\ref{difference_wkb}). As the branch cut is not visible in an expansion of this term in $X=\exp(x)$, we attribute the sign to $\phipert$ in the notation of section \ref{subsection: consequences qc}.

\subsection{Testing the quantization conjecture for complex $\hbar$} \label{subsection:numerics}
Much numerical evidence has been provided for the quantization condition (\ref{quantization_condition}) in the case of local $\IP^1 \times \IP^1$ for real $\hbar$ \cite{Grassi:2014zfa}. This in particular is the geometry that features prominently in the analysis of the ABJM matrix model in \cite{Marino:2009jd} and related works. In this section, we will extend this study to complex $\hbar$. Note that the operator $\cO_{\mirrorcurve}$ obtained from quantization of $P_{\mirrorcurve}$ in (\ref{mirrorcurvep1p1}) is invariant under the conjugation $\hbar \rightarrow \bar{\hbar}$, as is the quantization condition (\ref{quantization_condition}). Every study at complex $\hbar$ hence tests the quantization condition simultaneously inside and outside the $q$ unit circle.

To determine the eigenvalues numerically, we use the formula (\ref{harmonic_expectation}) to compute the matrix elements of the operator $\tcO_{\mirrorcurve}$ given in (\ref{spectralp1p1}) in the basis (\ref{basis_harm_osc}) of harmonic oscillator eigenstates up to a fixed level $n$, and then diagonalize the matrix numerically. The dependence on the choice of $\omega$ and $m$ in (\ref{basis_harm_osc}) decreases with increasing matrix size.

To evaluate the quantization condition, we first compute the refined topological string partition function on local  $\IP^1 \times \IP^1$ using the refined topological vertex \cite{Iqbal:2007ii,Taki:2007dh}. This computation is detailed in \cite{Hatsuda:2015qzx} for the general case of $A_n$ singularities fibered over $\IP^1$, and will not be reviewed here. The vertex formalism computes the series coefficients $a_n$ in
\be
\Ztop \sim_{Q_1} \sum_{n} a_n(q,t,Q_2) Q_1^n 
\ee
as rational function in the variables $Q_2$, $q= e^{i
	\epsilon_1}$, and $t=e^{- i \epsilon_2}$. The $\epsilon_1 \rightarrow -\epsilon_2$ limit reproduces the conventional topological string partition function, as computed in \cite{Iqbal:2003ix,Iqbal:2003zz}, with modulus $Q=Q_1$ and mass parameter $Q_m = Q_2 / Q_1$. The limit 
\be
\lim_{\epsilon_2 \rightarrow 0} \epsilon_2 \log \Ztop = \FNS
\ee
yields the NS limit of the topological string amplitude that enters into the quantization condition. The first few terms are given by
\be
\FNS \sim_{Q_1,Q_2}  \frac{q+1}{q-1} Q_2 - \frac{q (q+1)}{(q-1)(q-Q_2)(q Q_2 -1)} Q_1 + \ldots \,.
\ee
Here, the first term is the leading contribution in a series in $Q_2$ of $Q_1$ independent terms, and the second is the order 1 term in a $Q_1$ expansion.
 
The choice of $\hbar$ for which we can test the quantization condition must satisfy several constraints. As the quantization condition is implemented as a truncated series in $Q_i$ and $Q_i^{\frac{2 \pi}{\hbar}}$, we need to ensure that the solution to the quantization condition lies at sufficiently small values of $Q$ such that both expansion parameters are small. Also, values of $\hbar$ for which either $|e^{i \hbar}|$ or $|e^{\frac{4 \pi^2 i}{\hbar}}|$ are large (order 100 or more) lead to unstable numerics. 
 
\begin{figure}
	\centering
	\includegraphics[width=7cm]{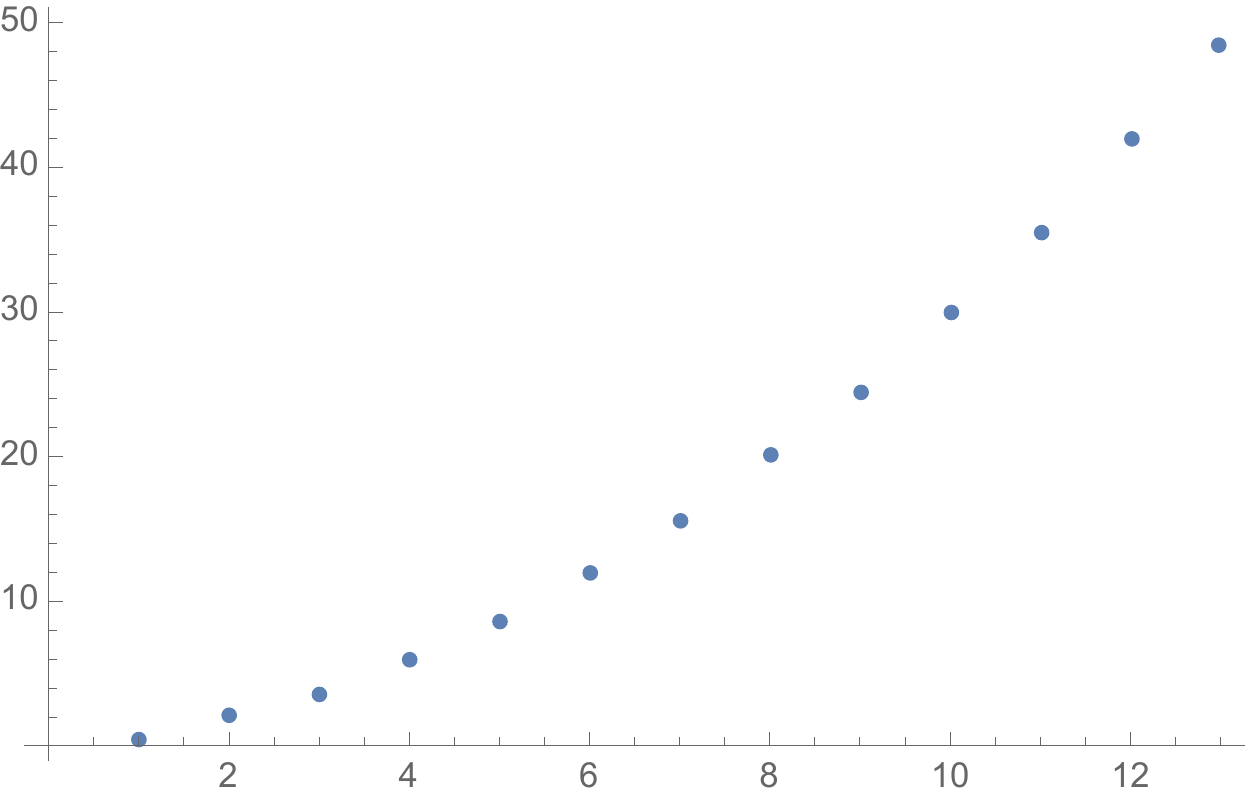}
	\caption{\small The largest coefficient $d_{n,k}$ plotted against $n$.} \label{figure_growth_FNS}
\end{figure}

We first consider the eigenvalue problem at $z_1=z_2$, i.e. $Q=Q_1=Q_2$, $Q_m=1$. Upon expansion of $\FNS$ as
\be  \label{fns_expansion}
\FNS \sim_Q \sum_n b_n(q) Q^n \,,
\ee
we find the coefficients $b_n(q)$ to be rational functions of the form
\be
b_n(q) = \frac{\sum c_{n,k} \cos(d_{n,k} \,\hbar)}{\sin(\frac{n}{2}\,\hbar)} \,, \quad c_{n,k} \ge 1 \,,
\ee
where $\max_{k} d_{n,k}$ grows faster than linearly in $n$, see figure \ref{figure_growth_FNS}.
It follows that for complex $\hbar$, $|b_n(q)|$ is unbounded and the series (\ref{fns_expansion}) that enters into the quantization condition does not converge. We see this behavior reflected in table \ref{table:non_convergent}, where we have evaluated the quantization condition at successive orders in $Q$. Had $\FNS$ been convergent, we would have expected an increasing number of digits of $z$ to stabilize with increasing order. Instead, we see that the result appears to stabilize to a certain number of digits, but then oscillates around this value. 

\begin{table}[h]
	\centering \tiny
	\begin{tabular}{| c | c |}
		\hline
		order in $Q$ &  $z$ via quantization condition \\
		\hline
		$ 1$ & $7.8881124 +  2.03196930 \,i$ \\
		$ 2$ & $8.1353674 +  1.96234286 \,i$ \\
		$ 3$ & $8.1238152 +  1.97363235 \,i$ \\
		$ 4$ & $8.1242759 +  1.97257505 \,i$ \\
		$ 5$ & $8.1242830 +  1.97266253 \,i$ \\
		$ 6$ & $8.1242794 +  1.97265470 \,i$ \\
		$ 7$ & $8.1242805 +  1.97265560 \,i$ \\
		$ 8$ & $8.1242800 +  1.97265549 \,i$ \\
		$ 9$ & $8.1242805 +  1.97265566 \,i$ \\
		$10$ & $8.1242795 +  1.97265546 \,i$ \\
		$11$ & $8.1242819 +  1.97265761 \,i$ \\
		$12$ & $8.1242664 +  1.97267005 \,i$ \\
		$13$ & $8.1242882 +  1.97280531 \,i$ \\
		\hline
	\end{tabular}
\caption{\small The quantization condition evaluated at $\hbar = 3+i$.}\label{table:non_convergent}
\end{table}
Never the less, the prediction of the quantization condition, evaluated at optimal truncation in the expansion in $Q$, reproduces the result obtained for $z$ via numerical diagonalization to numerous significant digits, see table \ref{table:numerics}.
 
For the examples that we consider, it turns out that the solutions of (\ref{quantization_condition}) for $Q$ at larger $n$, i.e. for higher lying eigenvalues, have smaller absolute value. This explains the improved accuracy of the results at larger $n$ in table \ref{table:numerics}. In fact, beyond $n=0$, the results via the quantization condition stabilize to more significant digits than those from numerical diagonalization up to matrix size 
$500 \times 500$.

\begin{table} 
	\centering \tiny
\begin{tabular}{| c | c | c | c |}
	\hline
	$\hbar$ & $n$ & $z$ via diagonalization & $z$ via quantization condition \\
	\hline
	      & 0 & $8.12428024641619 +  1.97265543644422 \,i$  & $8.124280 + 1.972655 \, i \,\,(9)$\\
	$3+i$ & 1 & $19.06647674202373 +  8.65025419938627 \,i$ & $19.06647674202373 + 8.65025419938627 \, i \,\, (12)$   \\
	    & 2 & $36.171976898401704 + 22.4710366010966 \,i$ & $36.171976898401704536102 + 22.4710366010966616226996 \, i \,\, (13)$ \\
	\hline
	          & 0 & $32.59048527302 + 24.768795735781 \,i$&  $32.590485 + 24.76879 \, i \,\,(5)$ \\
	$10+3\,i$ & 1 & $149.88891552236 + 180.10000255910 \, i$ & $149.88891552236089 + 180.100002559106430 \,i \,\, (7)$\\
	         & 2 & $429.46307908 + 757.2311848397 \, i$ & $429.46307908198397242150 + 757.23118483976591204907 \, i \,\, (7)$ \\
	\hline
\end{tabular}
\caption{\small Numerical diagonalization with matrix size $500 \times 500$, best approximation via quantization condition is given, with the order at which the approximation is attained indicated in parentheses When more digits stabilize up to the maximal order (13) considered via the quantization condition than via diagonalization, these are indicated, even though the stabilization will be lost at higher order.}\label{table:numerics}
\end{table}

We can also check the quantization condition away from the $Q_1=Q_2$ locus. To this end, we diagonalize the operator (\ref{cosh_form}) at a fixed value of $z_m$, and evaluate the quantization condition at $Q_2 = z_m Q_1$. Note that $\FNSBPS$ as determined by the refined vertex is exact in $Q_2$. The quantum mirror map however is only known in an expansion in this parameter. For consistency, we hence also expand $\FNSBPS$ in $Q_2$ before evaluation.

The results for two choices of $z_m$ are recorded in table \ref{table:numerics_Q12}.  

\begin{table}[h]
	\centering \tiny
	\begin{tabular}{| c | c | c | c |}
		\hline
		$z_m$ & $n$ & $z$ via diagonalization & $z$ via quantization condition \\
		\hline
		      & 0 & $6.55723612994535 + 7.20861330852542 \, i$ & $36.55723 + 7.20861 \, i$ (7) \\
		$10$  & 1 & $75.8984079656015 + 31.385294436428 \, i$  & $75.8984079656015 + 31.38529443642880 \, i$ (9) \\
		      & 2 & $137.8896673007909 + 80.706970674681 \, i$ & $137.889667300790980981 + 80.7069706746815527222 \, i$ (9) \\
		\hline      
			  & 0 & $21.21608102907488235 + 7.855443627204370422 \, i$ & $21.21608 + 7.85544 \,i $ (6)\\
		$5+i$ & 1 &	$45.53184055833938617 + 26.8836637285323608 \, i$ &	$45.531840558339 + 26.88366372853 \, i$ (8) \\
			  & 2 & $82.8551978931157334 + 64.5662794096170074 \, i$ & $82.855197893115733449488 + 64.56627940961700744288 (9) \, i$ \\
		\hline
	\end{tabular}
\caption{\small These results are obtained at $\hbar = 3+i$. Numerical diagonalization with matrix size $500 \times 500$. Same conventions as in table \ref{table:numerics}.}\label{table:numerics_Q12}
\end{table}

\newpage 
\section{Conclusions}
We have argued that the rules of exact WKB analysis carry over to difference equations, and used these to determine the monodromy behavior of WKB solutions. The quantization condition (\ref{quantization_condition}) then reduces to a question regarding the monodromy of the elements of the kernel of the quantized mirror curve $\cO_{\mirrorcurve}$. We have argued that the contribution non-perturbative in $\hbar$ to the quantization condition (\ref{quantization_condition}) proposed in \cite{Franco:2015rnr} arises when requiring that the kernel of the quantum mirror curve $\cO_{\mirrorcurve}$ have non-trivial intersection with a particular function space $\functionspace$ defined in section \ref{subsection:domain}.

The analysis performed in this paper should be enhanced in several directions:

To accumulate evidence for the exact WKB rules as applied to difference equations, or to discover their limitations, they should be tested in the case of difference equations with known exact solutions. 

The relation between the quantum $B$-period and $\FNS$ which enters centrally in the quantization condition (\ref{quantization_condition}) relies on the Nekrasov-Shatashvili conjecture (\ref{NS_conjecture}). It would be important to have a proof of this conjecture, perhaps along the lines of the proof in \cite{Kashani-Poor:2014mua} in the case of $\cN=2^*$ 4d gauge theory. This might help clarify the required $B$-field dependent shift in the quantum mirror map alluded to in footnote \ref{footnote:shift}.

The numerical manifestation of the quantization of the complex structure parameters $z_i$ in the higher genus case should be clarified.

We have emphasized the need of specifying the function space $\functionspace$ on which the equation $\cO_{\mirrorcurve} \Psi = 0$ is to be solved. The possibility has been raised in the literature that the Borel resummation of the naive WKB solution automatically imposes $\Psiw \in \functionspace$ \cite{Hatsuda:2015fxa}. In the case of real $\hbar$, evidence was presented in \cite{Grassi:2014cla} that the Borel-Pad\'e resummation of the $\hbar$ expansion of $\FNS$ on local $\IP^1 \times \IP^1$ at real $\hbar$ is smooth. See also \cite{Hatsuda:2015owa} for an analysis of the conifold geometry for real $\hbar$. This issue merits further study for general $\hbar$. 

The relationship between flat open coordinates and distinguished forms of the operator $\cO_{\mirrorcurve}$ should be further explored. Also, the correlation between the $B$-field required for the pole cancellation mechanism in the open and the closed case deserves further study. For a recent study linking open to closed string invariants, see \cite{Hatsuda:2016rmv}.

Very recently, an article \cite{Aganagic:2016jmx} appeared on the arXiv studying the monodromy of difference equations in very different language from that employed in this paper. It would be interesting to see how the two analyses are related.

\section*{Acknowledgments}
We would like to thank Jie Gu, Albrecht Klemm, and Marcos Marino for helpful conversations and correspondence, and Jie Gu for comments on the draft. 

We acknowledge support from the grant ANR-13-BS05-0001.

%

\bibliography{refs}
\bibliographystyle{utcaps}

\end{document}